\documentclass[journal,comsoc]{IEEEtran}
\usepackage[T1]{fontenc}

\ifCLASSINFOpdf
\else
\fi
\usepackage{mdwlist}

\usepackage{tabularx}
\usepackage{graphicx}
\usepackage{epstopdf}
\usepackage{algorithmic}
\usepackage{multirow}
\usepackage{subfig}
\usepackage{amsmath}
\begin{document}
%
\title{A Survey on Mobile Edge Networks: Convergence of Computing, Caching and Communications}
%
%
%
%
\author{Shuo~Wang,
        Xing~Zhang, \emph{Senior Member, IEEE},
        Yan Zhang, \emph{Senior Member, IEEE},
        Lin Wang,
        Juwo Yang,
        and~Wenbo~Wang, \emph{Senior Member, IEEE}
\thanks{S. Wang, X. Zhang, L. Wang, J. Yang and W. Wang are with the School
of Information and Communications Engineering, Beijing University of Posts and Telecommunications, Beijing 100876,
China (e-mail: wangsh@bupt.edu.cn; zhangx@ieee.org; wanglin121@gmail.com; yangjuwo@163.com; wbwang@bupt.edu.cn).}
\thanks{Y. Zhang is with University of Oslo, Norway (e-mail: yanzhang@ieee.org). }
\thanks{This work is supported by the National Science Foundation
of China (NSFC) under grant 61372114, 61571054
and 61631005, by the New Star in Science and Technology
of Beijing Municipal Science \& Technology Commission
(Beijing Nova Program: Z151100000315077). }}

\maketitle

\begin{abstract}

As the explosive growth of smart devices and the advent of many new applications, traffic volume has been growing exponentially. The traditional centralized network architecture cannot accommodate such user demands due to heavy burden on the backhaul links and long latency. Therefore, new architectures which bring network functions and contents to the network edge are proposed, i.e., mobile edge computing and caching.
Mobile edge networks provide cloud computing and caching capabilities at the edge of cellular networks. In this survey, we make an exhaustive review on the state-of-the-art research efforts on mobile edge networks. We first give an overview of mobile edge networks including definition, architecture and advantages. Next, a comprehensive survey of issues on computing, caching and communication techniques at the network edge is presented respectively. The applications and use cases of mobile edge networks are discussed. Subsequently, the key enablers of mobile edge networks such as cloud technology, SDN/NFV and smart devices are discussed. Finally, open research challenges and future directions are presented as well.
\end{abstract}


%
\IEEEpeerreviewmaketitle

\section{Introduction}

\IEEEPARstart{D}{uring} the past several decades, mobile cellular networks have been evolving steadily and significantly from the 1st generation (1G) voice only systems to current 4th generation (4G) all-IP based LTE-Advanced networks. The system capacity and average data rate have improved greatly with the technology advancements in physical layer such as WCDMA, OFDMA, MIMO, CoMP and in network layer such as heterogeneous network (HetNet) and cloud radio access network (C-RAN). According to a recent report from Cisco \cite{Cisco}, the mobile data traffic has grown 4000-fold during the past 10 years and will continue grow at a rate of 53 percent annually from 2015 to 2020. In particular, mobile video traffic accounts for more than half of total mobile data traffic and this percentage keeps increasing. Besides, mobile devices are getting smarter in their computing capabilities, and new machine type devices appear such as wearable devices and sensors in addition to human type devices. This leads to massive M2M connections in next generation mobile networks.

Machine type communications (MTC) bring a wide range of new applications and services in wireless networks. The authors in \cite{MTC stat} presented the current status and challenges of MTC for cellular systems. The most important challenges include massive number of MTC devices, small data bursts, low-latency, and low power consumption. Various solutions have been proposed to accommodate these challenges \cite{MTC arch}, \cite{MTC que}. Since the processing capabilities of MTC devices are constrained, one promising solution is to offload their tasks to places that have powerful processing capabilities. The ubiquitous connectivity of MTC leads to the strong heterogeneous networking paradigm. Research efforts have been made to accommodate such MTC applications from 4G to the emerging 5G systems \cite{IoT in 5G}.

\begin{figure}[!t]
\centering
\includegraphics[width=2.5in]{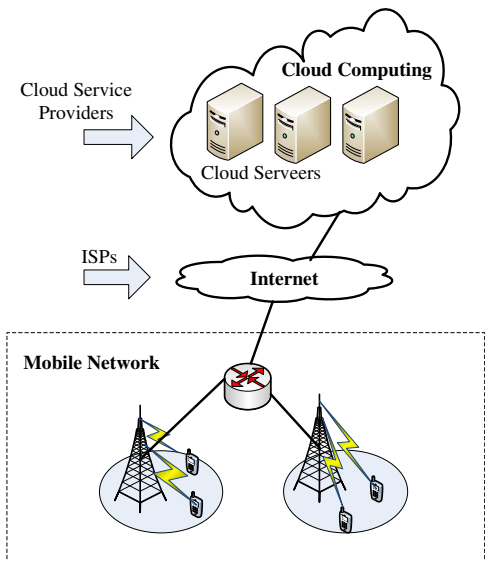}
\caption{Architecture of Mobile Cloud Computing.}
\label{MCC_arch}
\end{figure}

The preliminary mobile computing scheme adopted a 2-level hierarchy which originally called "servers" and "clients" \cite{MC}. Later on, The terminology "cloud" was used to represent a collection of servers with computational and information resources, which leads to the research on mobile cloud computing (MCC). Mobile cloud computing considers various mobile-related factors compared to the traditional computation offloading techniques, such as device energy, bandwidth utilization cost, network connectivity, mobility, context awareness and location awareness \cite{MCC app}, \cite{MCC Zhang}. Various survey articles have been published focusing on different aspects of MCC. In \cite{MCC_Guan} and \cite{MCC}, the authors presented generic issues on mobile cloud computing including architecture, technical challenges and applications. In \cite{MCC ZTE}, existing works on mobile cloud platforms and access schemes were discussed. The authors compared two mobile cloud platforms, the Hyrax platform \cite{Hyrax} and virtual machine (VM) based cloudlets \cite{cloudlet}, and then reviewed intelligent access schemes utilizing the user's location and context \cite{MCC access}. The authors in \cite{MCC app} elaborated the entities affecting computation offloading decision and presented detailed application models classification and the latest mobile cloud application models. The authors in \cite{MCC Fern} presented a detailed taxonomy of mobile cloud computing based on the key issues and the approaches to tackle them, such as operational issues, end user issues, service level issues, security, context-awareness and data management. User authentication is significant in securing cloud-based computing and communications. In \cite{MCC Auth}, the authors surveyed the state-of-the-art authentication mechanism in MCC and compare it with that in cloud computing. The merits of MCC can be summarized as follows. Firstly, it can provide sufficient resources for mobile devices and has great flexibility. Secondly, the cost of MCC can be reduced due to centralized management of resources. Finally, since all the tasks are processed in the cloud, MCC support multiple platforms. Fig. \ref{MCC_arch} illustrates the general architecture of MCC, which contains 2 tiers: the cloud and the mobile devices.

Despite the merits of mobile cloud computing, it faces the inevitable problems such as long latency and backhaul bandwidth limitation due to the long distance from mobile devices to the Internet cloud. Mobile Edge computing (MEC), which deploy cloud servers in base stations, is a promising solution for the problem since the computation capability is closer to the mobile devices \cite{MEC ETSI}. The benefits of MEC consist of low latency, proximity, high bandwidth, real time radio network information and location awareness. MEC is recognized as one of the key technologies for the next generation 5G networks by the European 5G PPP (5G Infrastructure Public Private Partnership) \cite{5G Vision}. The authors in \cite{MEC survey} described the taxonomy of MEC based on different aspects including its characteristics, access technologies, applications, objectives and so on. They also identified some of the open issues in MEC. In \cite{MEC taxo}, the authors classified several applications deployed at the mobile edge according to the technical metrics: power consumption, delay, bandwidth usage and scalability. The benefits of MEC each stakeholder can get were also analyzed. For mobile users and network operators, the main profit is reduced delay which leads to fast services, For the application service providers (ASP), they can benefit from the utilization of user-related information.

Fog computing is another edge paradigm which supports the future internet of things (IoT) applications \cite{Fog role in IoT}. It uses near-user edge devices such as edge routers to carry out substantial amount of computing tasks. Although it is a concept similar to MEC in some aspects, it distinguishes itself as more suitable for the context of IoT \cite{fog survey}. The OpenFog Consortium was founded by Cisco Systems, ARM Holdings, Dell, Intel, Microsoft and Princeton University in 2015 to promote development and interests in fog computing. The representative applications and various aspects of research issues of fog computing were highlighted in \cite{fog survey}. Three driving scenarios which will benefit from fog computing are augmented reality and real-time video analytics, content delivery and mobile big data analytics. The quality of service (QoS) metrics for fog services include four aspects: connectivity, reliability, capacity and delay. The interfacing and programming model, resource management, security and privacy are key challenges fog computing encounters. In \cite{fog sec}, the authors surveyed the new security and privacy challenges fog computing faces in addition to those same as cloud computing, such as secure data storage and secure computation. The security threats and challenges in the edge paradigms were analyzed in \cite{MEC sec}. The authors presented specific challenges and promising solutions in the following aspects: identity and authentication, access control system, protocol and network security, trust management, intrusion detection system, privacy and virtualization.

A similar edge computing concept proposed by the academia is called Cloudlet \cite{vm based cloudlet}, \cite{cloudlet}. The cloudlet is an extension of distant "cloud". It is a "data center in a box" which is self-managed, energy efficient and simple to deploy on a business premises such as a coffee shop or an office room. One approach for deploying cloudlet infrastructure is to integrate cloudlet and WiFi access point hardware into a single entity. However, the management of widespread deployed cloudlet is challenging. The solution is transient customization of cloudlet infrastructure using hardware VM (virtual machine) technology \cite{vm based cloudlet}.

The increasing demand for massive multimedia services over mobile cellular network poses great challenges on network capacity and backhaul links. The emergence of mobile edge caching and delivery techniques are promising solutions to cope with those challenges \cite{cache in air}. In traditional centralized mobile network architecture, content requests of end users are served by remote internet content providers. In that situation, duplicate traffic has to be transmitted through the whole mobile network, which leads to network congestion and a waste of network resources. Caching popular contents at the network edge (e.g., gateways, base stations and end user devices) can avoid duplicate transmissions of same content and improve users' quality of experiences due to reduced latency. The concept of caching is not new in cellular networks. It has been used in web caching and information centric networks. Ali et al. \cite{web caching} investigated the web caching and prefetching techniques in improving the web performance. A classification of caching policies was presented such as recency-based policies (e.g., LRU), frequency-based policies (e.g. LFU), size-based policies, function-based policies and randomized policies. Furthermore, the authors in \cite{web cache replace} described the advantages and disadvantages of cache replacement strategies and outlined potential research topics in modern proxy caches. The authors in \cite{Cache ICN} and \cite{Cache ICN EE} surveyed caching mechanisms in information centric networks and its energy efficiency of caching respectively. However, due to the characteristics of wireless cellular networks, the above mentioned caching techniques cannot be directly applied. The caching schemes in cellular networks should be thoroughly investigated.

Despite the merits of the above edge paradigms, they make the wireless network heterogeneous and difficult to manage in the traditional way. Emerging technologies such as network function virtualization (NFV) and software defined networks (SDN) are promising solutions which enable the network to be flexible and easy to maintain. Network function virtualization is a recently proposed network architecture that utilizes IT virtualization technologies to virtualize network node functions on top of standard general purpose hardware, which changes the communication network infrastructure to be more cost-efficient \cite{NFV vEPC}. Software defined network is a computing network architecture that split the function of control plane and data plane. In legacy mobile networks, both the control plane and user plane (or data plane) are integrated on the macro base stations. This network architecture cannot meet the explosive traffic and connection growth in future 5G networks. Based on the concept of SDN, new architectures are proposed. Zhang et al. \cite{DOC} proposed a macro-assisted data-only carrier scheme which split the control function and data function to macro base stations (MBSs) and small base stations (SBSs), respectively. Pentikousis et al. \cite{SDMN} introduced and validated a software-defined mobile network architecture which improved the capability of the operator and reduced time to market for new services. A summary of existing survey articles on mobile computing and caching is shown in Table \ref{tab:MCC}.

The above mentioned advantages and progress of mobile edge networks motivate us to perform a comprehensive literature survey.
The main contributions of this article are summarized as follows:
\begin{enumerate}
\item A comprehensive survey of mobile edge network architectures is presented, including MEC, Fog Computing, Cloudlets and edge caching. A comparison of different edge computing proposals is summarized. The advantages of mobile edge networks are pointed out.
\item A comprehensive survey of the key technologies in mobile edge computing and caching is presented. In particular, as for computing at the network edge, the literatures related to computation offloading, cooperation between the edge and core network, combination with 5G and proposed platforms are elaborated. As for edge caching, research progress on content popularity, caching policies, scheduling, and mobility management are surveyed.
\item The applications and use cases of mobile edge networks are comprehensively summarized. The key enablers of mobile edge networks are pointed out, including cloud technology, software defined network, network function virtualization and smarter mobile devices.
\item The open issues and challenges related to mobile edge networks are identified, such as network heterogeneity, realtime analytics, pricing, scalability, utilization of wireless big data, context awareness and so on.
\end{enumerate}

\begin{table}[!t]
\renewcommand{\arraystretch}{1.3}
\caption {SUMMARY OF EXISTING SURVEY ARTICLES ON MOBILE COMPUTING AND CACHING}
\label{tab:MCC}

\centering

\begin{tabular}{|p{1.6cm}| p{0.9cm}|p{4.5cm}|}
\hline
\textbf{Aspects} & \textbf{Survey Papers} & \textbf{Contributions} \\
\hline
             & \cite{MCC app} & A classification of application models and investigation of latest mobile cloud application models\\
             \cline{2-3}
Mobile Cloud Computing & \cite{MCC_Guan} & A summary of challenges for MCC, application partition and offloading technologies, classification of contexts and context management methods.\\ \cline{2-3}
             & \cite{MCC} & An overview of MCC definition, architecture, and applications, as well as the generic issues and exi
             sting solutions. Discussions of the future research directions of MCC.\\
             \cline{2-3}
             & \cite{MCC ZTE} & An investigation of existing works on representative platforms and intelligent access schemes of MCC.\\ \cline{2-3}
             & \cite{MCC Fern} &  A detailed taxonomy of mobile cloud computing based on the key issues and the approaches to tackle them.\\ \cline{2-3}
             & \cite{MCC Auth} & A comprehensive survey of the state-of-the-art authentication mechanism in MCC and comparison with that in cloud computing.\\
\hline

Mobile Edge Computing & \cite{MEC survey} & A taxonomy of MEC based on different aspects including its characteristics, access technologies, applications, objectives and so on. Identification of some of the open issues in MEC.\\ \cline{2-3}
                & \cite{MEC taxo}&  A classification of applications deployed at the mobile edge according to the technical metrics and the benefits of MEC for stakeholders in the network.\\ \cline{2-3}
                & \cite{MEC sec} & A discussion of the security threats and challenges in the edge paradigms, as well as the promising solution for each specific challenge. \\

\hline

Fog Computing & \cite{fog survey}& Highlighting the representative applications and various aspects of research issues of fog computing.\\ \cline{2-3}
               & \cite{fog sec}& Survey of the new security and privacy challenges fog computing faces in addition to those same as cloud computing. \\
\hline
Caching     &\cite{web caching} & An investigation of the web caching and prefetching techniques in improving the web performance as well as a classification of caching policies.\\ \cline{2-3}
             &\cite{web cache replace}  & A description of the advantages and disadvantages of cache replacement strategies, outlining potential research topics in modern proxy caches.\\\cline{2-3}

            &\cite{Cache ICN} & A survey of caching mechanisms in information centric networks.       \\ \cline{2-3}
             &\cite{Cache ICN EE} & A survey of the energy efficiency of caching in information centric networks.  \\

\hline

\end{tabular}
\end{table}

The rest of the paper is organized as follows. In Section~\ref{sec:overview}, an overview of mobile edge networks including the definition, architecture and advantages is presented. In Section~\ref{sec:mec}, an elaborated survey of literatures on computing related issues at mobile edge networks is given. In Section~\ref{sec:cache}, the research efforts on edge caching are fully surveyed. The advances in communication techniques with synergy of computing and caching is discussed in Section \ref{sec:comm}.
In Section~\ref{sec:apps}, The applications and use cases of mobile edge networks are explained. In Section~\ref{sec:enabler}, the key enabler technologies of mobile edge networks are summarized. The open challenges and future directions are shown in Section~\ref{sec:issue}. Finally, conclusions are drawn in Section~\ref{sec:conc}.
For convenience, a summary of all abbreviations is shown in Table \ref{tab:abbr}.

\begin{table}[!t]
\renewcommand{\arraystretch}{1.3}
\caption {SUMMARY OF ABBREVIATIONS}
\label{tab:abbr}

\centering
\begin{tabular}{l l}
\hline
5G & 5th generation\\
5GPPP & 5G Infrastructure Public Private Partnership\\
API & application interface \\
AR & augmented reality\\
ASE & area spectral efficiency\\
ASP & application service provider\\
BS & base station\\
C-RAN & cloud radio access network\\
CAPEX & capital expenditures \\
CDN & content delivery network \\
CoMP & coordinated multiple point\\
CSI & channel state information\\
D2D & device-to-device\\
DC & data center\\
EPC & evolved packet core\\
ETSI &  European Telecommunications Standards Institute\\
HetNet & heterogeneous network\\
IA & interference alignment\\
ICN & information centric network \\
IoT & internet of things \\
IRM & independent reference model \\
ISP & internet service provider\\
IP & internet protocol\\
LFU & least frequently used\\
LRU & least recently used \\

LTE & long term evolution\\
M2M & machine to machine \\
MBS & macro base station \\

MCC & mobile cloud computing\\
MEC & mobile edge computing\\
MEN & mobile edge networks\\
MIMO & multiple in multiple out\\
MPV & most popular video \\

MTC & machine type communication\\
Multi-RAT & multiple radio access technology \\
NFV & network function vitualization\\
OFDMA & orthogonal frequency division multiple access\\
OPEX & operating expenses \\
QoE & quality of experience \\
RACS & radio application cloud servers \\
RAN & radio access network\\
RNC & radio network controller\\
RTT & round trip time\\
SBS & small base station \\
SDN & software defined network\\
SINR & signal to interference plus noise ratio \\
SNM & shot noise model \\
SVC & scalable video coding\\
TDMA & time division multiple access  \\
UGC & user generated content \\
UPP & user preference profile \\

VM & virtual machine\\
VR & virtual reality\\
WCDMA & wideband code division multiple access\\

\hline
\end{tabular}
\end{table}

\section{Overview of Mobile Edge Networks}
\label{sec:overview}

The evolvement of mobile cellular networks has experienced 4 generations in the last two decades with the advancements in information and telecommunications technology. At the same time, users' demands for mobile networks also become more and more strict such as ultra high data rate and extremely low latency. Moreover, various new requirements are appearing due to the advent of new kinds of smart devices and new applications, such as virtual reality and the internet of things. The traditional base station centric network architecture cannot fulfill these requirements any more. The mobile cellular network architecture is evolving from BS-centric to device centric \cite{5G survey} and content centric network in the future 5G system, where the center of gravity moves from the network core to the edge \cite{5G dir}. In this section, we will firstly explain what is mobile edge networks. Then the architecture of mobile edge networks is presented. At last, we discuss the advantages of mobile edge networks.

\subsection{What is Mobile Edge Networks}
The core idea of mobile edge networks is to move network functions, contents and resources closer to end users, i.e., the network edge, by utilizing SDN and NFV technologies. The network resources mainly include computing, storage or caching, and communication resources. While in some literature caching is included in computing resources \cite{MEC taxo}, in this paper we discuss them separately since the service types and problems they are related to are different.

Edge computing in mobile networks is evolved from mobile cloud computing, which is an architecture moving the computing power and data storage away from mobile devices and into the cloud to leverage the powerful computing and storage capability of cloud platform \cite{MCC}. However, mobile cloud computing faces several challenges such as long latency and high backhaul bandwidth consumption, therefore it is not suitable for real-time applications. The authors in \cite{MCC} list the technical challenges that MCC faces in detail. In the mobile communication side, the challenges include low bandwidth, service availability and heterogeneity due to the characteristics of wireless networks such as scarce radio resources, traffic congestion, and multiple radio access technologies (multi-RAT). In the computing side, efficient and dynamic computing offloading under environment changes is challenging, as well as the security issues for users and data, efficiency of data access and context awareness. In comparison, edge computing enables the network edge to have cloud computing capabilities.

Three different edge computing schemes have been proposed by industrial and academic parties: mobile edge computing, fog computing and cloudlet. Mobile edge computing was proposed by the standards organization European Telecommunications Standards Institute (ETSI) \cite{MEC ETSI}. It is based on a virtualized platform which enables applications running at the network edge. Meanwhile, the infrastructure of NFV can be reused by applications which is beneficial for network operators. MEC servers can be deployed at various location at the network edge such as the LTE macro base station (eNodeB), at the 3G Radio Network Controller (RNC), and at an aggregation point. The deployment location may be affected by scalability, physical constraints, performance criteria and so on. MEC applications can be seamless deployed on different MEC platform intelligently and flexibly based on the technical parameters such as latency, required resources, availability, scalability and cost. The objective of ETSI MEC is to deliver a standard architecture and industry-standardized APIs for 3rd party aplications \cite{Fog mec cloudlet}.

Fog computing is another edge computing architecture with the aim to accommodate IoT applications originally proposed by Cisco \cite{Fog mec cloudlet}. Fog computing is an extension of the cloud computing paradigm to the wireless network edge \cite{Fog role in IoT}. The name \emph{fog computing} comes from the analogy that the fog is closer to people than the clouds. Similarly, the distance of an IoT device is closer to a fog computing platform than to lage-scale data centers. The necessity of fog computing maintained by Cisco is that a 2-tiered deployment of IoT applications is not sufficient for the requirements of low latency, mobility, and location awareness \cite{Fog mec cloudlet}. The solution is a multi-tiered architecture which deploys an intermediate fog platform between the device and the main cloud. The main characteristic of fog computing is that is is a completely distributed, multi-layer cloud computing architecture where the fog nodes are deployed in different network tiers \cite{fog iot}.

The concept of Cloudlet is developed by an academic team at Carnegie Mellon University \cite{vm based cloudlet}. It can be deployed in both Wi-Fi networks and cellular networks. The key features of cloudlets are near-real-time provisioning of applications to edge nodes and handoff of virtual machine images among edge nodes when a device moves \cite{Fog mec cloudlet}.

Mobile edge caching is proposed for tackling the challenges of massive content delivery in future mobile networks. The advances in storage enable the network to exploit the large amount of low cost storage resources at different places in the network. The traffic load of cellular network is dynamically varying in the spatial and temporal domain \cite{ST model}. Proactive caching is an approach that exploits such traffic dynamicity by proactively cache popular content during off-peak periods, which reduces peak traffic demands \cite{living on edge}. Since the cache units are deployed at the network edge, a lot of information can be exploited to improve the caching efficiency. For example, social structures of users can be leveraged to cache and disseminate content via D2D communications.

Based on the above discussion, we define mobile edge networks as:
\emph{"A mobile network architecture that deploys and utilizes flexible computing and storage resources at the mobile network edge, including the radio access network, edge routers, gateways and mobile devices etc., with the help of SDN and NFV technologies"}.

\subsection{Architecture of Mobile Edge Networks}
The mobile edge networks introduce new way of manipulating computing and storage resources. Both industries and the academia bring up proposals on the architecture of MEN. We will present the specific architectures including ETSI MEC, Fog Computing, Cloudlet and Edge Caching. Then we will summarize these proposals and give the general architecture of MEN.

\subsubsection{Mobile Edge Computing}
\begin{figure}[!t]
\centering
\includegraphics[width=2.5in]{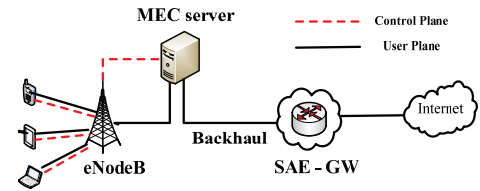}
\caption{Architecture of MEC.}
\label{MEC}
\end{figure}

Mobile edge computing has drawn much attention of industries and the academia. In industries, the ETSI has launched an Industry Specification Group (ISG) on MEC in December 2014. The ISG produces specifications what enable the hosting of 3rd-party innovative applications in a standard MEC environment \cite{MEC ETSI}. The group has delivered several specifications on service scenarios, requirements, architecture and APIs. Fig. \ref{MEC} shows the architecture of MEC. MEC servers are located in proximity of base stations. They can either handle a user request and respond directly to the UE or forward the request to remote data centers and content distribution networks (CDNs) \cite{MEC white paper}.

\subsubsection{Fog Computing}
\begin{figure}[!t]
\centering
\includegraphics[width=2.5in]{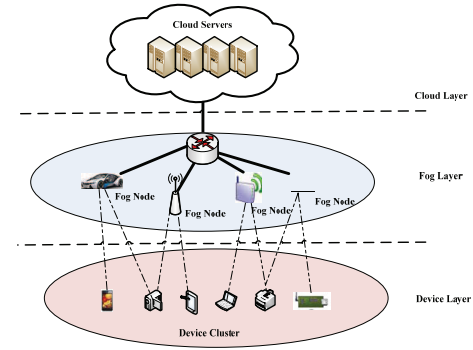}
\caption{Architecture of Fog Computing.}
\label{Fog}
\end{figure}

Fog computing is a platform designed mainly for Internet of Things use cases. Its component fog nodes are massively distributed in wide area. The main feature of fog is that it utilizes collaborations among multiple end user clients or near-user edge devices to help processing and storage of mobiled devices \cite{Fog network}. Compared to Cloud, Fog has advantages in three dimensions: exploiting storage, computing and control functions, communication and networking at or near the end user \cite{Fog IoT Overview}. In the view of fog computing, the edge is part of the core network and a data center. Fog and Cloud complement each other to make computing, storage and communication possible anywhere along the continuum between the cloud and endpoints. Fog computing is also integrated to the C-RAN architecture to formulate the Fog RAN architecture \cite{Fog RAN}.
The architecture of fog computing is shown in Fig. \ref{Fog}. It contains three layers: cloud layer, fog layer and device layer. The Fog layer may contain multiple tiers according to the requirement. The Fog node could be small BSs, vehicles, WiFi Access Point and even user terminals. The devices choose the most appropriate fog node to associate with.

\subsubsection{Cloudlet}
\begin{figure}[!t]
\centering
\includegraphics[width=2.5in]{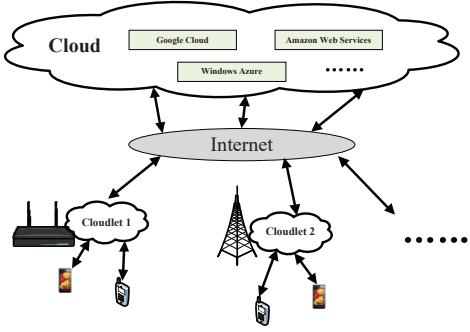}
\caption{Architecture of Cloudlets.}
\label{cloudlet}
\end{figure}

The cloudlet proposal is a 3-tier architecture: "device - cloudlet - cloud" \cite{Fog mec cloudlet}. Cloudlets could be deployed at WiFi access points or LTE base stations \cite{cloudlet}. Many new applications require end to end latency of 1 ms. Theoretically, 1 ms of propagation delay requires a cloudlet within 300 km even at the speed of light, In reality, cloudlets should be deployed much closer to ensure the delay requirement. The combination of 5G cellular networks and cloudlets will make this possible \cite{cloudlet eco}. Fig. \ref{cloudlet} illustrate the general architecture of Cloudlet systems. To overcome the limited capabilities of single cloudlet, cooperation among different cloudlets is necessary in order to meet the user demands \cite{MEC cloudlet}. A comparison of different mobile computing architectures is summarized in Table \ref{tab:comp_MC}.

\begin{table*}[!t]
\renewcommand{\arraystretch}{1.3}
\caption{COMPARISON OF DIFFERENT MOBILE COMPUTING ARCHITECTURES}
\label{tab:comp_MC}

\centering
\begin{tabular}{|p{2.5cm}|p{2.5cm}|p{2.5cm}|p{2.5cm}|p{2.5cm}|}
\hline
\textbf{Item}  & \textbf{MCC} & \textbf{MEC} & \textbf{Fog Computing} & \textbf{Cloudlet}\\
\hline
Originally proposed by     & Not specific  & ETSI & Cisco  & Prof. Satyanarayanan\\
\hline
Hierarchy     & 2 tiers  & 3 tiers  & 3 or more tiers & 3 tiers\\
\hline
Latency & High & Low & Low & Low \\
Ownership & Centralized by cloud providers: Amazon, Microsoft, etc.
& Mobile operators & Decentralized Fog node Owners & Local business\\

\hline
Sharing Population & large & Medium & Small & Small\\

\hline
Location & Large data center & RAN & between devices and DC & between devices and DC, or directly in a device\\
\hline
Context awareness & No & Yes & Yes & Could be\\
\hline
Cooperation between nodes & No & No & Yes & No \\
\hline

\end{tabular}
\end{table*}

\subsubsection{Edge Caching}
\begin{figure}[!t]
\centering
\includegraphics[width=2.5in]{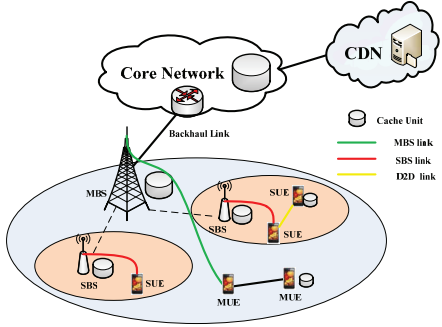}
\caption{Architecture of Edge Caching.}
\label{cache}
\end{figure}

Caching in the mobile edge network has been proved beneficial. The future mobile networks will be heterogeneous due to dense deployment of different types of base stations. Thus, cache can be deployed at various places in the mobile networks. In legacy cellular system, the content requested by users has to be fetched from the Internet CDN node far away from the mobile networks. Then, caching content at the mobile core network is implemented. However, the backhaul links are still constrained. In addition, with the evolvement of base station and low cost storage unit, deploying cache at macro base stations and small base stations become feasible. In the future 5G networks, D2D communication enables the storage unit at user devices to be exploited for content sharing according to the social relations among users. A general architecture of edge caching is shown in Fig. \ref{cache}.

\begin{figure*}[!t]
\centering
\includegraphics{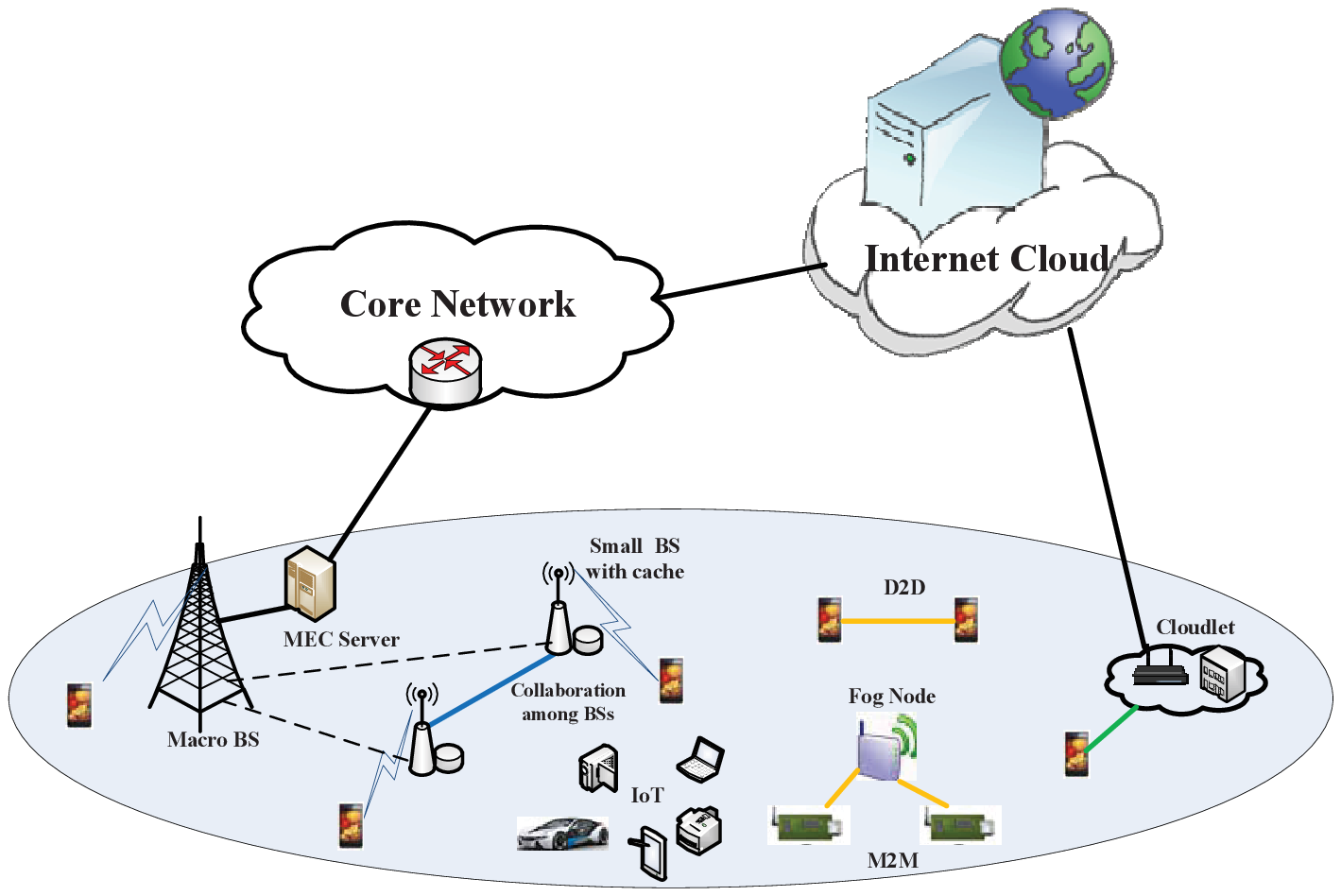}
\caption{Architecture of Mobile Edge Networks.}
\label{MEN}
\end{figure*}

Based on the above discussion, the general architecture of mobile edge networks is shown in Fig. \ref{MEN}.

\subsection{Advantages of Mobile Edge Networks}
Compared to traditional centralized network architecture, researchers found the mobile edge networks have advantages in various aspects. Next we will discuss them in detail.

\subsubsection{Reduced Latency}
Since the processing and storage capabilities are in proximity to the end users, the communication delay can be reduced significantly. The main applications which benefit from this are computational offloading and video content delivery. The work of \cite{MEC Rimal} shows the offload packet delay can be reduce without affecting the network performance using a solution that combine MEC and cloud. The authors in \cite{mec quantify} confirm that offloading through edge computing platforms can achieve great improvement of latency for highly interactive and compute intensive application such as augmented reality and cognitive assistance, in both WiFi and LTE networks. Gao et al. \cite{cloudlet} present experimental results from Wi-Fi and 4G LTE networks showing that offloading to cloudlets can improve response times by 51\% compared to cloud offloading.
\subsubsection{Bandwidth Reduction}
The deployment of edge servers in mobile edge network infrastructure can save operation cost by up to 67\% for the bandwidth-hungry applications and compute intensive applications \cite{MEN benefit}. Research results show that backhaul savings can be up to 22\% exploiting proactive caching scheme\cite{living on edge}. Higher gains are possible if the storage capability is increased.
\subsubsection{High Energy Efficiency}
The experiment results show that edge computing can save energy consumption significantly for different applications in both WiFi networks and LTE networks \cite{mec quantify}. The energy consumption of applications using central data centers (DCs) in cloud computing and those using nano data centers in fog computing is compared in \cite{Fog Energy}. Results show that using nano DCs can save energy, which is affected by the following factors: type of access network, ratio of active time to idle time of nano DCs and type of applications. It is found that the energy consumption in a mobile device can be reduced by up to 42\% through offloading to cloudlets compared to cloud offloading \cite{cloudlet}.

\subsubsection{Proximity Services}
The architecture of mobile edge networks has great advantages in providing proximity services since the edge servers are closer to end users and D2D communication technology can be exploited \cite{MEC Nunna}. Therefore, the traffic load on radio access network can be reduced.

\subsubsection{Utilization of Context Information}
The MEC server deployed at different places in the radio access network can obtain detailed context information including network level information, device level information \cite{MEC Nunna}. With this information, the resources of network can be allocated more efficiently and the user experience can be improved. For example, location based applications can be deployed directly on MEC server rather than on the Internet cloud far from the users.

\section{Computing at Mobile Edge Networks }
\label{sec:mec}
Computing is a major resource in mobile networks. Many compute intensive applications are appearing recently such as augmented reality, high definition video streaming and interactive gaming. However, the computation capability is very constrained in mobile devices. In addition, the power consumption of these computing tasks is very high for the battery capacity of current mobile devices. Edge computing paradigm enables the possibility of offloading the computation tasks to more powerful edge servers. How beneficial is computing at the edge networks? Many works have been done to find the answer to this question. In this section, we will survey the state-of-the-art research efforts on this issue.

\subsection{Objectives}
The benefits of edge computing is various. Different applications or system may have different performance requirements. We will present some of the common performance objectives that edge computing can obtain.
\subsubsection{Minimize Energy Consumption}
Many works have been done to evaluate the energy efficiency of edge computing. Various optimization schemes have been proposed to minimize energy consumption in both the network side and the device side. For computation offloading in 5G heterogeneous networks, the energy cost both the task computing and file transmission should be considered. Zhang et al. \cite{MEC in 5G} design an energy efficient computation offloading scheme, which jointly optimizes radio resource allocation and offloading to minimize the energy consumption of the offloading system under the latency constraint. In their scheme, the devices are classified into three types according to their ability and requirements firstly. Then they allocate the wireless channels of MBSs and SBSs to mobile devices according to their priority iteratively until the radio resources are used up or all the devices have been allocated required channels. At each iteration step, the scheme ensures the system obtains minimum energy cost. The results show that the proposed scheme has lower energy consumption that computing without offloading especially  with large number of mobile devices. In \cite{ennergy Fog}, the authors investigate the energy aware interaction between fog and cloud. It is demonstrated that the overall energy consumption of data centers can be reduced without a significant deterioration of the network performance.

\subsubsection{Maximize Capacity}
The next generation 5G networks require support of 1000 times higher mobile data volume per area than current 4G LTE networks \cite{metis 5G}. This requires more capacity in the RAN, backhaul and fronthaul. Offloading is one of the combination of technologies that address these challenges in the RAN in addition to more spectrum, network densification and higher spectrum efficiency \cite{5G considerations}. The strategy that combine the fog and cloud operations can achieve high system capacity while providing low latency for requested services \cite{Fog Cloud Souza}.
\subsubsection{Minimize Latency}
Latency is an important performance metric that affects user experience. The latency requirements of next generation 5G networks is 1 ms round trip time (RTT), which is almost 10 times reduction from the 10 ms RTT in 4G \cite{5G survey}. For real-time applications, the delay incurred by offloading tasks to the cloud is unacceptable. Enhancing the high density SBS with computing capabilities is a much more feasible way. The authors in \cite{MEC power control} propose a distributed cloud-aware power control algorithm that is suitable for delay sensitive applications. In \cite{MEC delay opt}, a delay minimization problem is formulated under the constraint of power consumption. The authors design an optimal computation task scheduling policy for MEC systems. The delay of general traffic flows in the LTE downlink can be minimized by service level scheduling via MEC server deployed at the eNodeB \cite{MEC min delay}. Fog also provides low delay but with low capacity limitation, while the combined operation of fog and cloud can minimize the service latency and guarantee the capacity requirements at the same time \cite{Fog Cloud Souza}.

\subsection{Computation Offloading}
One of the main purposes of edge computing is computation offloading to break the limitations of mobile devices such as computational capabilities, battery resources and storage availability. When and how to offload the computation tasks is a hard problem. Various approaches have been proposed to tackle this problem under many kinds of scenarios such as single user case, multi user case and in vehicular networks \cite{MEC Mao}, \cite{multiuser clustering}, \cite{MEC Zhang}. Moreover, in next generation heterogeneous networks, the computation tasks can not only be offloaded to servers but also to devices by utilizing D2D communication \cite{MEC in 5G}, \cite{femtoclouds}. A summary of literatures on computation offloading is shown in Table \ref{tab:computation off}.

\subsubsection{Single User Case}
For the single user case, the optimal selection of executing mobile applications in the mobile device (mobile execution) or offloading to the cloud need to be analyzed \cite{MCC Zhang}. One of the common design objective is to save energy for the mobile device. Consider the stochastic channel condition in wireless networks, the optimal scheduling policies should be obtained. In \cite{MEC Zhang}, a threshold based scheduling policy is derived which depends on the energy consumption model and wireless channel model. In \cite{MEC Mao}, a low complexity online algorithm is proposed for a MEC system with energy harvesting devices. The algorithm is called the Lyapunov optimization-based dynamic computation offloading algorithm which jointly determines the offloading decision, the CPU-cycle frequencies and the transmit power for offloading.
The authors in \cite{MEC Partial offloading} investigate the partial computation offloading problem with the aims of minimizing device energy consumption and latency of application execution. They consider both the single server scenario and multiple servers scenario. The results show the conditions under which local execution is optimal and conclude that total offloading is not the optimal choice when the device has the capability of dynamic voltage scaling.
\subsubsection{Multi-user Case}
The Offloading problem in multi-user case is more complex than that in single user case. Many research efforts have been made on this topic. Chen et al. \cite{MEC Chen} show that the multi-user computation offloading problem is NP-hard and propose a distributed game theoretic approach for efficient computation offloading decision. The authors in \cite{MEC Sard} jointly optimize the radio resources and computational resources to minimize the overall users' energy consumption with latency constraints in a MIMO multicell system where multiple users offload computing tasks to a common cloud server. An iterative algorithm based on successive convex approximation technique is proposed to solve the nonconvex problem. Huang et al. \cite{MEC Huang} study the optimal resource allocation for a multi user mobile edge computation offloading system with both time division multiple access (TDMA) and orthogonal frequency division multiple access (OFDMA). The authors in \cite{MEC Kety} form the multiuser computation offloading as a multiple knapsack problem and propose a heuristic algorithm to solve it. In \cite{MEC  sequen}, the inter-cell interference environment of dense small cells and limitation of computation resources of MEC are considered. An adaptive sequential offloading game approach is proposed which adaptively adjust the number of offloaded users to reduce the queuing delay. The authors in \cite{MEC Opportunistic} propose an opportunistic computation offloading scheme for online data stream tasks. The scheme is evaluated using the event data stream of 5 million activities collected from 12 users for 15 days and results show significant data reduction by 98\%.

\subsubsection{Offloaded to MEC Server}
The common place to offload computation tasks is MEC servers. The REPLISOM architecture proposed in \cite{REPLISOM} offload the memory objects produced by multiple IoT devices to the edge cloud located at the LTE eNodeB. It reduces the latency and cost during offloading. In \cite{ME VoLTE}, the video encoding process during video call is offloaded to the MEC edge server based on a communication protocol for negotiating the offloading strategy, which reduces energy consumption of mobile devices during video call. In \cite{Load distribution}, the issue of load balancing in multiuser fog computing scenario is addressed and the authors propose a low complexity algorithm for fog clustering. A two layer radio access points clustering algorithm is proposed which achieve better quality of experience (QoE) than the centralized and decentralized only strategies \cite{multiuser clustering}. In \cite{MEC task offloading}, the authors take into consideration the energy consumption, time delay and execution unit cost to the design the optimal scheme for task offloading. They define a concept of opportunity consumption based on the limited energy and computing units of a mobile device.

\subsubsection{Offloaded to devices}
With the technology advancement of smart devices, more computation resources can be leveraged using D2D technology. A collection of co-located mobile devices can be utilized to provide cloud services at the edge \cite{femtoclouds}. In this scenario, computation tasks can be offloaded to other mobile devices nearby other than to the MEC server. The task scheduling problem should be reinvestigated which is different from that of offloading to servers.

\subsubsection{Mobility Awareness}
The user mobility cannot be ignored in mobile edge networks. Due to user mobility, the contact time between users and MEC servers is dynamic which will impact the strategy of offloading, specifically, where and what to offload \cite{Mobility Chen}. The authors in \cite{mobility model} propose a mobility model which suggests that the inter-contact time between any two users complies with an exponential distribution. With this model, the mobility aware computation offloading strategy is developed in \cite{Mobility Chen}.

\begin{table*}[!t]
\renewcommand{\arraystretch}{1.3}
\caption{SUMMARY OF LITERATURES ON COMPUTATION OFFLOADING}
\label{tab:computation off}

\centering
\begin{tabular}{|p{3 cm}|p{3 cm}|p{8 cm}|}
\hline
\textbf{Work Area}  & \textbf{Related Work} & \textbf{Key Points} \\
\hline
Single User System    & \cite{MCC Zhang}, \cite{MEC Mao}. \cite{MEC Partial offloading}   &
\begin{itemize}
\item Computation offloading choice
\item optimal scheduling policies
\item partial offloading problem
\end{itemize}\\
\hline
Multi User System    &  \cite{MEC Chen}, \cite{MEC Sard}, \cite{MEC Huang}, \cite{MEC Kety}, \cite{MEC  sequen} &
\begin{itemize}
\item Multi-user computation offloading problem is NP-hard
\item Distributed game theoretic approach
\item Joint optimization of radio and computational resources
\item Heuristic optimization algorithm
\item Minimize energy consumption and reduce delay
\end{itemize} \\
\hline

Offloaded to MEC Server    &   \cite{REPLISOM}, \cite{ME VoLTE}, \cite{Load distribution}, \cite{multiuser clustering}  &
\begin{itemize}
\item Offload objects produced by multiple IoT devices to MEC server
\item Reduce latency, energy consumption and execution cost
\item Load balancing
\item AP clustering algorithm
\end{itemize}\\
\hline
Offloaded to Devices    &  \cite{femtoclouds}  &
\begin{itemize}
\item Using D2D technology
\item A collection of co-located mobile devices to provide cloud services
\item task scheduling problem
\end{itemize}\\
\hline
Mobility Awareness    & \cite{Mobility Chen}, \cite{mobility model}   &
\begin{itemize}
\item dynamic contact time between users and MEC servers
\item Mobility model
\item mobility aware offloading strategy
\end{itemize}\\

\hline
\end{tabular}
\end{table*}

\subsection{Cooperation between the edge and the core}
Despite its advantages of mobile edge networks such as low latency and high energy efficiency, the computing resources of edge networks are still limited. To exploit the merits of both the edge networks and the cloud platforms at the core network, the cooperation between them is valuable. The workload allocation problem is studied in a combined fog-cloud computing system in \cite{Fog cloud}. The authors design an approximation solution to obtain the tradeoff between power consumption and delay. Simulation results show that cooperation between fog and cloud can significantly improve the performance of cloud computing. A hybrid approach combining fog computing and public/private IoT clouds is proposed in \cite{edge assisted cloud} to enable integrated IoT applications. The proposed solution can improve the number of task requests being successfully executed.
The authors in \cite{Fog Cloud Souza} combine the operation of fog and cloud to simultaneously improve the latency performance and system capacity.

\subsection{Platforms}
In addition to the theoretical discussions of mobile edge computing, some preliminary implementation of MEC servers are developed. Nokia Neworks introduced a real-world MEC platform in 2014: the Radio Application Cloud Servers (RACS) \cite{RACS}. RACS use a VM hypervisor to deploy virtual machine images which execute MEC applications. In \cite{Fog Industrial IoT}, an adaptive operation platform (AOP) is proposed to manage fog computing infrastructure based on the operational requirements of industrial IoT process. The AOP includes several functional elements: the Model Building element, the Rule Mapper element and the Rule Deployer element. The authors in \cite{Fog platorm D2D} exemplify the M2M Fog platform and evaluate three different deployment to connect Fog nodes to the LTE network: the legacy deployment based on macro cells, D2D based and small cell based deployment.

\section{Caching at Mobile Edge Networks}
\label{sec:cache}
To exploit caching technology in the mobile networks, many issues should be studied. Where to cache? What to cache? How to cache? The traditional caching schemes in content centric networks have not consider the wireless networks characteristics such as dynamic traffic load and interference \cite{cache dynamic traffic}. In this section, we will survey the research efforts that have been made in the mobile edge networks. The related issues include caching places, content description, caching policies, content delivery and so on. A summary of literatures on edge caching is shown in Table \ref{tab:caching}.

\subsection{Caching Places}
There are many places where caching units can be deployed in the mobile networks. In the all-IP based cellular networks, three main places that can deploy cache are the core network, the radio access network (RAN) and user devices. Currently the widely deployed places of caching is the evolved packet core (EPC) \cite{cache in air}. By caching content at the mobile core network, the mobile traffic can be reduced by one to two thirds. Moreover, deploying caching at the EPC is technically easier than at the RAN. The caching places at the edge networks are discussed as follows.

\subsubsection{MBS Caching}
In heterogeneous networks, MBSs have more coverage areas and can serve more users. Caching at MBS can obtain better cache hit probability. In \cite{video aware scheduling}, the performance of reactive and proactive caching at MBS is studied. A video aware backhaul and wireless channel scheduling technique is proposed combining with edge caching. The results demonstrate that the video capacity can be significantly increased and the stalling probability of videos is reduced. The authors in \cite{Proactive Storage} investigate the  storage allocation problem in MBS caching. They propose a heuristic method to solve the NP-hard problem.

\subsubsection{SBS Caching}
SBSs are densely deployed in next generation heterogeneous networks. Therefore, caching at SBSs is another good choice since the SBSs are more close to the end users and usually provide higher data rate. Many literatures have studied the performance of caching at SBSs \cite{caching modeling bustug}, \cite{learning based caching}, \cite{femtocaching}, \cite{multicast cache}, \cite{caching and routing}.
\subsubsection{Device Caching}
D2D communication is one of the key technologies in next generation 5G networks. The storage resources in mobile devices can be exploited. The QoE of users can be greatly improved by caching contents in mobile devices if the caching strategy is carefully designed. A caching based D2D communication scheme is proposed in \cite{D2D caching social} considering the social relations among users and their common interests. In \cite{D2D caching Yang}, an opportunistic cooperative D2D transmission scheme exploiting caching capability at the devices is proposed. In this scheme, the D2D users are divided into clusters and different popular files are cached at the users within a cluster. Results show that the proposed strategy can provide 5 to 6 times throughput gain over existing D2D caching scheme when the popularity distribution is skewed.

\subsection{Content Popularity}
To decide what to cache in the edge networks, the popularity of content should be considered to maximize the hit probability of cache, i.e., the probability that the content request by users is cached in the edge networks.
\subsubsection{Static Model}
Most of the current works on mobile caching assume that the content popularity is static and adopt an independent reference model (IRM): the request of contents is generated according to an independent Poisson process whose rate is related to content popularity modeled by a power law \cite{cache misconceptions}. The commonly used popularity model is the zipf model observed in web caching \cite{Zipf}.
\subsubsection{Dynamic Model}
The static IRM model cannot reflect the real content popularity which is time varying \cite{cache misconceptions}. A dynamic popularity model called the shot noise model (SNM) is proposed in \cite{SNM}. The model uses a pulse with two parameters to model each content: the duration reflects the content life span and the height reflects its instantaneous popularity. In \cite{video popularity}, the authors analyze the statistical properties of user generated content (UGC) popularity distributions and discuss opportunities to leverage the" long tail" video demands.

\subsection{Caching Policies and Algorithms}
Various Caching policies and algorithms have been proposed in mobile caching. Some of the conventional caching policies in wired networks are revised for wireless networks. In addition, new schemes such as learning based policies and cooperative caching policies are also proposed. The literature \cite{Caching Policy survey} reviews in detail the conventional caching policies and forwarding mechanism in information centric networks. we will present a taxonomy of the caching policies in wireless mobile networks in the following, which is summarized in Table \ref{tab:caching policy}.

\subsubsection{Conventional Caching Policies}
Content replacement policies such as the least frequently used (LFU) and least recently used (LRU) have been adopted in a large number of caching policies \cite{Caching Policy survey}, \cite{LRU}. These strategies are simple and efficient with uniform size objects. However, these policies ignore the  download latency and size of objects. Another proactive caching policy used in content deliver networks is the MPV policy, which caches the most popular videos based on the global video popularity distribution \cite{video aware scheduling}. However, the cache size of the RAN is very limited compared to that of CDN. The hit probability achieved by MPV policy could be too low for RAN caches.
\begin{table*}[!t]
\renewcommand{\arraystretch}{1.3}
\caption{SUMMARY OF LITERATURES ON EDGE CACHING}
\label{tab:caching}

\centering
\begin{tabular}{|p{3 cm}|p{3 cm}|p{8 cm}|}
\hline
\textbf{Work Area}  & \textbf{Related Work} & \textbf{Key Points} \\
\hline
Content Popularity   & \cite{cache misconceptions}, \cite{Zipf}. \cite{SNM}, \cite{video popularity}   &
\begin{itemize}
\item Static IRM model, power law distribution
\item Dynamic model, SNM model
\end{itemize}\\
\hline
Caching Policies and Algorithms   &  \cite{Caching Policy survey}, \cite{video aware scheduling}, \cite{Learning caching}, \cite{cache replace}, \cite{cooperative caching and deli}  &
\begin{itemize}
\item Conventional caching policies
\item User preference based polices
\item Learning based policies
\item Non-cooperative policies
\item Cooperative policies
\end{itemize} \\
\hline


Caching File Types   &  \cite{video aware scheduling}, \cite{video popularity}, \cite{cache SVC},  \cite{IoT Caching} &
\begin{itemize}
\item Multimedia data
\item IoT data
\end{itemize}\\
\hline
Mobility Awareness    & \cite{mobility method}, \cite{MobiCacher}, \cite{Mobility on hit}, \cite{exploit mobility}, \cite{mobility prediction}, \cite{Edge caching mobility predic}    &
\begin{itemize}
\item Spatial and temporal properties of user mobility
\item Discrete time Markov chain model
\item Short connection duration problem in mmWave 5G networks
\end{itemize}\\

\hline
\end{tabular}
\end{table*}

\subsubsection{User Preference Based Policies}
In \cite{video aware scheduling}, the authors propose a user preference profile (UPP) based caching policy. It is observed that local video popularity is significantly different from national video popularity and users may show strong preferences toward specific video categories. The UPP of each user is defined as the probability that a user requests videos of a specific video category.

\subsubsection{Learning Based Policies}
In fact, the content popularity is time-varying and is not known in advance. Therefore, the track and estimation of timely content popularity is an important issue. Based on machine learning technology, learning based caching polices are proposed in \cite{Learning caching}. The authors in \cite{Learning caching} solve the problem of distributed caching in SBSs from a reinforcement learning view. By adopting coded caching, the caching problem is reduced to a linear program that considering the network connectivity and the coded aching scheme performs better than the uncoded scheme.
The authors in \cite{cache replace} solve the cache replacement problem with a Q-learning based strategy.

\subsubsection{Non-Cooperative Caching}
Some of the existing caching policies decide the content to cache at each base station without consider the cooperation among BSs. In \cite{video aware scheduling}, the proposed scheme makes caching decision based on the UPP of active uses in a specific cell without considering the impact of caches in other cells. In \cite{cache replace}, the cache replacement problem modeled as Markov Decision Process (MDP) is solved in a distributed way using Q-learning method, without exchanging extra information about cached data between the BS. This strategy outperforms the conventional ones such as the LFU, LRU and randomized strategy.

\begin{table*}[!t]
\renewcommand{\arraystretch}{1.3}
\caption{SUMMARY OF LITERATURES ON CACHING POLICIES AND ALGORITHMS}
\label{tab:caching policy}

\centering
\begin{tabular}{|p{3 cm}|p{3 cm}|p{8 cm}|}
\hline
\textbf{Work Area}  & \textbf{Related Work} & \textbf{Key Points} \\
\hline
Conventional Policies   & \cite{Caching Policy survey}, \cite{LRU}  &
\begin{itemize}
\item Least frequently used (LFU)
\item Least recently used (LRU)
\item most popular video (MPV)
\end{itemize}\\
\hline
User Preference Based Policies   &   \cite{video aware scheduling}   &
\begin{itemize}
\item Considering local content popularity
\item User's preference toward specific video categories
\end{itemize} \\
\hline

Learning Based Policies    & \cite{Learning caching}, \cite{cache replace}   &
\begin{itemize}
\item Estimation of timely content popularity with reinforcement learning
\item Q-learning based cache replacement strategy
\item Coded caching scheme

\end{itemize}\\
\hline
Non-cooperative Polices   &  \cite{video aware scheduling}, \cite{cache replace} &
\begin{itemize}
\item Distributed caching neglecting caches in other cells
\item Cache replacement problem modeled by MDP
\end{itemize}\\
\hline
Cooperative Policies    & \cite{Caching Distri}, \cite{cooperative caching and deli}, \cite{cache SVC}, \cite{cache redundancy}, \cite{D2D caching Yang}, \cite{caching and routing}, \cite{network coding}   &
\begin{itemize}
\item Cooperative cache management among base stations, UEs
\item Minimize bandwidth cost
\item Maximize traffic served from cache
\item Collaborative video caching and scheduling among  cells
\item Tradeoff between caching redundancy and diversity
\item Joint caching and routing design
\end{itemize}\\

\hline
\end{tabular}
\end{table*}
\subsubsection{Cooperative Caching}
A lot of existing works have studied the cooperation among cache places when designing the caching policies. In \cite{Caching Distri}, a light-weight cooperative cache management algorithm is developed to maximize the traffic volume served from cache and minimize the bandwidth cost. In \cite{cooperative caching and deli}, the cooperation between femto base stations and user equipment (UE) for content caching and delivery is investigated. The cooperative caching problem is formulated as an integer-linear programming problem and solved by using the subgradient method. The content delivery policy is formulated as an unbalanced assignment problem and solved by using Hungrian algorithm. The authors in \cite{cache SVC} explore the exploiting of scalable video coding (SVC) technique in collaborative video caching and scheduling among cells to further improve the the caching capacity and the QoE of users. In \cite{cache redundancy}, the authors investigate the cooperation among caches in the RAN and obtain the optimal redundancy ratio of content cached in each base station. The cooperation among D2D users is exploited for cache enabled D2D communication in \cite{D2D caching Yang}. A network coding based content placement method is proposed in \cite{network coding}. The strategy increases the amount of available data to users and results in a fair distribution of data at the same time. The authors in \cite{caching and routing} jointly design the caching and routing scheme to maximize the content requests served by small cell base stations under the constraint of BS bandwidth. The problem is reduced to a facility location problem and is solved using bounded approximation algorithms.

\subsection{Caching of Different File Types}
The most common file types for caching is multimedia files such as popular videos and audio files. The internet of things is one of the main use cases of the next generation 5G networks. Thus the caching of IoT data is also important as the IoT data volume is increasing and IoT data have different characteristics compared to multimedia data.
\subsubsection{Large Data Files or Multimedia Data}
The caching design of large multimedia data especially video files is studied in most of the existing literatures \cite{video aware scheduling}, \cite{video popularity}, \cite{cache SVC}. The characteristic of multimedia data is that many users have the same interest for popular videos. Therefore, caching the popular files in the RAN can benefit from high hit ratio.

\subsubsection{IoT Data}
The low-rate monitoring, measurement, and automation data generated by IoT applications ruuning on billions of devices need to be cached to reduce total traffic load. However, IoT data are different from multimedia data in that the IoT data have short lifetimes.  Thus, different caching policies are required. In \cite{IoT Caching}, the authors propose a model that takes into consideration both the communication costs and freshness of a transient IoT data item. The network load can be reduced significantly especially for highly requested data.

\subsection{Mobility Awareness}
The user mobility is a unique feature of wireless network, thus it should be considered in caching at the network edge. Many works have been done on this issue. The authors in \cite{mobility method} propose a general framework for mobility aware caching in content centric wireless networks. Both the spatial and temporal properties of user mobility are modeled. In \cite{MobiCacher}, the mobility pattern of users is considered and the mobility aware caching problem is formulated as an optimization problem aiming at maximize the caching utility and the authors propose a polynomial-time heuristic solution to solve the problem. The impact of user mobility on the hit performance of edge caching is analyzed in \cite{Mobility on hit}.
The user mobility is modeled as a discrete time Markov chain in \cite{exploit mobility}. The authors consider a scenario where segments of encoded content files are stored in a set of base stations in a cell with a main base station in it. The caching algorithm is designed to minimize the probability of using main base station for file delivery.
Different from the prior works that assume user mobility is known, the continuity of content service for the handover users is investigated based on mobility prediction algorithm when user mobility is unknown in \cite{mobility prediction}. In \cite{Edge caching mobility predic}, the authors exploit the information centric networking and the mobile Follow-Me Cloud approach to enhance the migration of content-caches located at the mobile network edge. The proposed content relocation algorithm improves content-availability by up to 500\% compared with existing solutions.

\subsection{Impact on System Performance}

\subsubsection{Capacity}
Existing works on edge caching have proved that caching at the network edge can significantly improve system capacity. For example, the solution proposed in \cite{video aware scheduling} can improve capacity by 3 times compared to having no cache in the RAN.

\subsubsection{Delay}
Caching at the network edge can significantly reduce content delivery delay due to the proximity of caches to end users. In \cite{video aware scheduling}, the initial delay and stalls of a video session is reduced by jointly scheduling RAN backhaul and wireless channels, therefore the video QoE is improved. In \cite{caching latency tradeoff}, the tradeoff between delivery latency and fronthaul and aching resources is derived.

\subsubsection{Spectral Efficiency}
In \cite{Hetnet caching}, the authors compare a cache-enabled 2-tier HetNet with a conventional HetNet without cache using stochastic geometry theory. The numerical results show that the cache-enabled helper density can be reduced by 3/4 compared to the pico BS density without cache to achieve the same area spectral efficiency (ASE). Given the total capacity within an area, an optimal cache enabled SBS density exists that maximizes the ASE.
\subsubsection{Energy Efficiency}
Energy Efficiency is another important performance metric for the next generation 5G networks. In \cite{caching energy Yang}, the impact of caching at BSs on the energy efficiency of downlink networks is analyzed. The results show that the energy efficiency will be improved when the file catalog size is small and caching at multiple pico BSs is more energy efficient than caching at a macro BS. In \cite{cache SD Hetnet}, the fundamental tradeoffs between energy efficiency and small cell density in software defined heterogeneous networks is demonstrated. The energy efficiency of cache enabled heterogeneous networks is much higher than current LTE networks.

\section{Advances in Communications Techniques with Synergy of Computing and Caching}
\label{sec:comm}
The combination of computing and caching resources with communication systems will change the design philosophy of communication networks to a large extent. Combining 5G with MEC would make certain inter and intra domain use cases feasible such as automotive services and e-Health services \cite{MEC Nunna}. By coordinating the cloud enabled small cells in a cluster, a distributed light data center can be created inside the communication networks \cite{MEC 5G}, \cite{MEC in 5G}. In this section, we will discuss the communication techniques and the synergy of communication, computing and caching.
\subsection{mmWave Communication}
The use of mm-wave spectrum in future networks enables high data rate transmission. Various schemes of integrating 5G BSs with legacy cellular networks have been proposed such as standalone mm-wave systems and hybrid systems of mm-wave BS and 4G BS \cite{5G survey}. The mm-wave spectrum is mainly used for data communications. The narrow beams used by mm-wave BSs can improve the link quality between BSs and large number of users. However, this kind of deployment also raise new challenges of short connection durations and frequent handoffs for high mobility users, making video streaming suffer from long latency. The authors in \cite{caching mmW} propose a caching-based mm-wave framework that precaches contents at the BS for handoff users. The proposed solution can provide consistent high quality video streaming for high mobility users in 5G mm-wave small cells.

\subsection{D2D Communication}
D2D communication enables the direct transmission between end devices in proximity. It has been recognized as one of the main technologies for 5G networks. The advantages of using D2D communication include one-hop communication, higher spectral efficiency, low transmission power, coverage extension and spectrum reuse \cite{D2D LTE}, \cite{D2D design aspects}. In addition, with the help of D2D communication, massive MTC devices can be offloaded from MBSs to SBSs, which improves overall network capacity and avoids traffic congestion at MBSs \cite{D2D offloading}. Many previous research efforts on D2D communication have been focus on the issues of spectrum reuse, peer discovery, power control, connection establishment and interference management \cite{D2D 3GPP}, \cite{D2D SC}. With the concept of edge caching and computing, the utilization of computing and storage resources on large number of smart devices draws the attention of researchers. Computation offloading and content sharing problems via D2D communications are under investigation by many researchers \cite{femtoclouds}, \cite{D2D caching social}, \cite{D2D caching Yang}.

\subsection{Transmission Schemes}
The ubiquitously deployed low-cost caches promote the development of content centric transmission. These caches at the mobile networks move the content near end users. However, the improvement of the effectiveness of small caches in small cells is a challenge. Since the content cached is popular for the local users, some requests for the same file may happen at nearby time. Therefore, these requests can be served via a single multicast transmission \cite{multicast cache}. Compared to caching schemes using unicast transmission, the multicast-aware scheme can reduce cost by up to 88\%. When the requested contents are all different in some time slots, a interweaved transmission scheme can be adopted to balance the traffic of sequential time slots of each SBS \cite{IA}.

\subsection{Interference Management}
In future ultra dense heterogeneous networks, interference is a key challenge. One of the potential technique for interference management of small cell networks is interference alignment (IA). In \cite{IA}, the authors investigate the IA problem of content centric communication with caching and computing exploited. Utilizing the content centric principle via caching, the topology of interference network is simplified and interference management becomes easier. Thanks to the high computational capacity of cloud computing platform, the solution of IA can be easily calculated. The proposed framework reduces backhaul load and the overhead of CSI feedback, and improve the throughput at the same time.
\subsection{Communication Resources Allocation and Scheduling}
With the integration of caching and computing resources in communication systems, the allocation and scheduling of communication resources is different from the legacy  networks. In addition to cache placement, the scheduling of communication resources also affect the efficiency of caching. In \cite{video aware scheduling}, joint design of video-aware backhaul scheduling and caching can improve capacity by more than 50\% than conventional policies. The careful design of wireless channel scheduling can result in significantly higher video capacity when the wireless channel bandwidth is constrained.  The joint caching and video scheduling strategy that combines collaborative caching with SVC is studied in \cite{cache SVC}.

\subsection{Synergy of Communication, Computing and Caching}
By enabling the mobile network with more computing capability, the scarce communication resources could be saved. In \cite{VRE}, a content-slimming system is proposed, which detect redundant video content and clip the from the original frames by computing and only transmit the necessary video content. This scheme reduce the transmission bandwidth consumption at least by 50\% compared to H.264 without sacrifice video quality and visual experience.

The application types, user mobility and communication resources will influence the optimal position of deploying computing and caching resources \cite{synergy}. For the low bandwidth, high persistence use cases that require computing tasks, computing resources should not be deployed at femtocells because frequent handovers and low computational power of the femtocells, especially when the backhaul transmission delays are minimal. For the high-bandwidth location-bound services require storage resources, such as AR, the computing and caching resources should be deployed as close to the end user as possible. In summary, computing and caching resources deployed at different layers of the network should be utilized according to the service types with the consideration of backhaul capacity.

\section{Applications and Use Cases}
\label{sec:apps}
The new applications are the main driven force for the evolvement of network architecture. The requirements of emerging applications become more and more strict in data rate, latency, etc. In this section, we will summarize the applications and use cases in mobile edge networks as shown in Table \ref{tab:application}.

\subsection{Dynamic Content Delivery}
With the increasing demand of multimedia content, the backhaul links face congestion problems in conventional centralized network architecture. Caching at the edge networks can provide dynamic content delivery based on the information of network status and user's context-aware information \cite{MEC survey}, \cite{web Zhu}. Since the content is placed close to mobile users, the QoE of mobile users is improved significantly.

\subsection{Augmented Reality/Virtual Reality}
The augmented reality (AR) and virtual reality (VR) technology is seen as the most promising application that will change our life style. This application needs real-time information of user's status such as the position and direction they are facing. The MEC server is capable of exploiting local context information and has powerful processing ability, which is very suitable for the AP/VR applications \cite{MEC scenarios}.

\subsection{Intensive computation assistance}
The computational capability is often sacrificed in order to lower the cost of devices. Therefore, computation offloading requiring very low processing time with low latency is necessary for such applications with intensive computation tasks. MEC servers are equipped  with high computational capabilities and can process the offloaded computation tasks in very short timescales. Moreover, the MEC server can collect information from multiple sources, which helps those devices perform tasks that require information from multiple sources \cite{MEC scenarios}.

\subsection{Video Streaming and Analysis}
It is observed that video traffic accounts for more than half of the total mobile data traffic in current networks and the percentage is still increasing. The adoption of edge caching avoids numerous redundant video streams transport through the core network to Internet CDNs. The use of MEC server allow video analysis at the more capable cloud platforms at network edge other than at the source producing videos \cite{MEC scenarios}, \cite{streaming}.

\subsection{Internet of Things}
The idea of Internet of Things is becoming a reality with the technology advancements in smart sensors, communications, and Internet protocols \cite{IoT survey}. Fog computing is an advantageous architecture in support of IoT applications. In \cite{Fog Industrial IoT}, the authors propose an adaptive operations platform which enable the application of Fog Component in industrial IoT context. The VM-based cloudlets at mobile edge networks enable edge analytics for crowd-sourced video content in IoT \cite{Edge Analytics}. MEC will provide new IoT services that would not be feasible before \cite{MEC Sabella}, \cite{MEC Corcoran}. The specific applications and use cases of MEN in the Internet of Things is presented below.

\subsubsection{Healthcare}
Real-time processing and event response are important for healthcare applications. Experiments proved that the healthcare system utilizing fog computing responded faster and was more energy efficient than cloud-only approaches \cite{Fog IoT Potential}. For example, fog computing can be used to detect falls of stroke patients.

\subsubsection{Wireless Sensor Systems}
Various scenarios that make use of wireless sensor systems can be use cases of mobile edge cloud computing platforms, such as in oil \& gas industry, building industry and environmental monitoring \cite{Fog mec cloudlet}.

\subsubsection{Smart Grid}
The analysis of the data generated in the smart grid environment is a very challenging task due to the complex parameters. The using of mobile edge computing can improve performance in throughput, response time and transmission delay \cite{MEC smart grid}. Smart Grid is a typical use case that require the interplay between the fog and the cloud \cite{Fog role in IoT}. The fog collect and process the local data generated by grid sensors and devices. The cloud provides global coverage and restore data that have a life cycle of months and years.
\subsubsection{Smart Home}
Smart home systems have become a new trend for future housewide ecosystems. Smart home is a kind of small-scale IoT system with limited spatial occupancy and localized communications. Deploying MEC servers as IoT gateways close to the smart objects will enable direct M2M interactions in future networks \cite{MEC smart home}. MEC node, which could be deployed on femtocells, home routers, set-top boxes and smartphones, is beneficial for low latency, localized and plug-and play services for smart home.

\subsubsection{Smart City}
The vision of smart city is to improve quality of life by utilizing technological advancements. The concept of MEC is helpful for time-critical events because the contents generated by large number of connected devices can be exploited to discover the occurrence of anomalous events \cite{MEC smart city}. Another useful component of smart city system is smart traffic lights. For instance, the smart traffic lights can send warning signals to approaching vehicles, or detect the pedestrians and cyclists who is crossing the street, or warn the vehicles of risks of running a red light \cite{Fog mec cloudlet}.

\subsection{Connected Vehicles}
Edge computing approaches can play an important role in connected vehicles, V2X communication and automotive safety services such as  real-time warning of ice on motorway and coordinated lane change maneuvers \cite{Fog mec cloudlet}. The applications run on MEC servers are in close proximity to the vehicles and can provide roadside functionality with low latency \cite{MEC Sabella}.
Traffic control and smart parking can be achieved since the edge network is able to collect and analyze real-time data from sensor devices installed ubiquitously \cite{MEC survey}.

\subsection{Cognitive Assistance}
Cognitive assistance applications are used to augment human perception and cognition ability. The authors in \cite{cloudlet cognitive} demonstrate how cloudlet can help in that cloudlet is only one wireless hop away cloud proxy. Therefore, it is the ideal offload site selection for cognitive assistance. All the latency-sensitive processing tasks are offloaded to cloudlet which the user device associates to. When the user moves away from the proximity of one cloudlet, the use will be handed off to another cloudlet in its proximity.
\subsection{Wireless Big Data Analysis}
Big data is commonly characterized along three dimensions: volume, velocity and variety. The authors in \cite{fog iot} suggests geo-distribution as a fourth dimension to the characterization of big data. For example, the large number of sensors and actuators are naturally distributed. Fog computing, as a distributed intelligent platform that manages distributed networking, compute and storage resources, is a promising choice to handle these big data. Compared to big data analytics performed at the core network, doing big data analytics at the network edge will reduce bandwidth consumption and network latency \cite{MEC survey}.

\begin{table*}[!t]
\renewcommand{\arraystretch}{1.3}
\caption{APPLICATIONS AND USE CASES OF MOBILE EDGE NETWORKS}
\label{tab:application}

\centering
\begin{tabular}{|p{4cm}|p{3cm}|p{7cm}|}
\hline
\textbf{Applications and Use Cases}  &\textbf{Related Work} &\textbf{Key Points} \\
\hline
Dynamic Content Delivery    & \cite{MEC survey}, \cite{web Zhu} &
\begin{itemize}
\item Placing content close to users
\item Exploiting user's context information
\end{itemize}\\
\hline
AR/VR     & \cite{MEC scenarios} &
\begin{itemize}
\item Real-time fast processing
\item Context aware
\end{itemize}
\\
\hline
Intensive Computation Assistance & \cite{MEC scenarios} &
\begin{itemize}
\item Low latency, low cost devices
\item Collecting info. from multiple sources
\end{itemize}\\
\hline
Video Streaming and Analysis & \cite{MEC scenarios}, \cite{streaming} &
\begin{itemize}
\item Avoiding redundant video streams transmission
\item More capable of analysis
\end{itemize}\\

\hline
Internet of Thing & \cite{Fog role in IoT}, \cite{Fog mec cloudlet}, \cite{Fog Industrial IoT}, \cite{IoT survey}, \cite{Edge Analytics}, \cite{MEC Sabella}, \cite{MEC Corcoran}, \cite{Fog IoT Potential}, \cite{MEC smart grid}, \cite{MEC smart home}, \cite{MEC smart city}&
\begin{itemize}
\item Healthcare
\item Wireless sensor systems
\item Smart grid
\item Smart Home
\item Smart City
\end{itemize}
\\

\hline
Connected Vehicles & \cite{MEC survey}, \cite{Fog mec cloudlet}, \cite{MEC Sabella} &
\begin{itemize}
\item V2X communication
\item Automotive safety services
\item Traffic control and smart parking
\end{itemize} \\
\hline
Cognitive Assistance & \cite{cloudlet cognitive}&
\begin{itemize}
\item Augmenting human perception and cognition ability
\item Processing latency sensitive tasks
\end{itemize} \\
\hline
Wireless Big Data Analysis & \cite{MEC survey}, \cite{fog iot}&
\begin{itemize}
\item reduce bandwidth consumption and network latency

\end{itemize} \\
\hline

\end{tabular}
\end{table*}
\section{Key Enablers}
\label{sec:enabler}
In this section, the key technologies that enable the concept of mobile edge networks become reality are presented. These technologies provide flexibility, scalability and operating efficiency to the mobile edge networks.
\subsection{Cloud Technology}
The concept of mobile edge networks is an extension of cloud computing capabilities to the edge of mobile networks. The advancement in cloud technology makes it easier to deploy virtual machines on high volume general purpose Servers in places such as base stations and gateways \cite{MEC survey}. The cloud can offer powerful processing capabilities and huge amount of resources. The integration of cloud and IoT has been proved beneficial for delivering new services \cite{Cloud IoT}. The mobile edge networks are integrated with cloud computing capabilities and offer effective solutions for service management and provision.

\subsection{Software Defined Network}
SDN technology enable the network to be intelligent, programmable, and more open \cite{AT&T}. The main idea of SDN is to separate the control and data planes. The benefits of SDN include creating network control planes on common hardware, exposing network capabilities through APIs, remotely controlling network equipment, and logically decoupling network intelligence into different software-based controllers \cite{AT&T}. SDN technology overcomes the shortcoming of management complexity due to large scale deployments of servers and applications \cite{MEC SDMEC}. Applying the SDN paradigm will enable management at different levels needed in the MEC platforms \cite{MEC Salman}.

\subsection{Network Function Virtualization}
Network function virtualization (NFV) is a complementary technology of SDN proposed for future 5G networks. The purpose of NFV is to virtualize a set of network functions by moving them from dedicated hardware to general purpose computing platforms using software technologies, which can provide the same services as legacy mobile networks. As a result, the capability of managing vast heterogeneous devices is improved as well as the scalability and flexibility of the network \cite{IoT in 5G}. With NFV, both the capital expenditures (CAPEX) and operating expenses (OPEX) of network operators are likely to be reduced. Network virtualization leads to the problem of virtual network embedding (VNE), VNE algorithms are currently being studied \cite{VNE}. The application of NFV changes the landscape of telecommunications industry and brings many benefits such as reducing the time to market, optimizing network configuration and topology in near real time, and supporting multi-tenancy \cite{ETSI NFV}.



\subsection{Smarter Mobile Devices}
In the legacy generations of cellular networks (2G, 3G, 4G), the system design is oriented by having complete control at the network infrastructure side. However, with the mobile devices become more powerful and smarter, this design philosophy should be changed to utilize intelligence at the device side \cite{5G dir}. In the future networks, the devices will play a more active role with more smartness. One important technology is the D2D communication. In current network, data traffic has become the main traffic types instead of voices. There are many situations where devices in close proximity would like to share content or interact with each other, for example, gaming and social networking. The direct D2D communication can improve network efficiency in several aspects. Firstly, it saves a lot of signaling resources and reduces transmission latency. Secondly, it can save large amount of energy compared to transmission through the help of base stations. Furthermore, the spectral efficiency can be improved since the path losses are much lower than BS to device communication. The D2D communication is expected to be natively supported in 5G network.
Another useful technology is local caching, which is intended to strike a balance between data storage and data transfer. For wireless devices, the marginal cost of transferring information is non-negligible. With more and more memory units installed in today's mobile devices, caching popular contents such as video or audio files at devices is clearly cheaper and more efficient than to transmit such content repeatedly via unicast since the demand is asynchronous.

\section{Open Challenges and Future Directions}
\label{sec:issue}
Mobile edge network is an revolution in the architecture of wireless mobile networks. It has many new features comparing to the existing 3G/4G cellular systems with better QoE performance and flexibility. Therefore, a wide variety of research challenges and opportunities exists for future research works. In this section, we point out the major open challenges in mobile edge networks and shed light on the possible future research directions.

\subsection{Open Research Challenges}
The stringent requirements of the next generation mobile networks, such as ultra high throughput, extremely low latency and high energy efficiency, pose great challenges on current research of both the academia and industry. We summarize the key research issues related to mobile edge networks below:
\subsubsection{Heterogeneity}
In the future networks, with the development of IoT and noval applications, the heterogeneity in networking, communication, and devices becomes a critical issue. This heterogeneity causes other related problems such as the asynchronism and non orthogonality \cite{5G survey}. The challenge of handling this heterogeneity under a unified network architecture need to be fully investigated.
\subsubsection{Computation Modeling}
To validate the accuracy of analysis and simulation works, the model of computation resources should be accurate. In current literatures, the computation resources are modeled as CPU cycles per second \cite{MEC Huang}. Although this model is simple for analysis, whether it fully reflects the characteristics of computing still remains a question. Hence, more accurate computation models should be developed.



\subsubsection{Enabling Realtime Analytics}
A lot of novel applications require realtime analytics such as VR/AR and e-Health. The dynamic resource management should be determined which schedules the analytic tasks to the most appropriate edge server guaranteeing the latency and throughput \cite{Fog IoT Potential}.

\subsubsection{User Mobility}
User mobility is a key challenge in mobile edge networks. It has non-negligible impact on caching and computation offloading decisions. The frequent mobility of users will cause frequent handovers among edge servers, Mobility management technique considering both horizontal and vertical mobility should be implemented allowing users to access edge servers seamlessly \cite{MEC survey}.
\subsubsection{Pricing Policy}
In mobile edge networks, the storage, computing and communication resources are allocated dynamically according to users' demand. Thus, the optimal pricing policy is different from the legacy systems. From a commercial perspective, the profits of all the stakeholders in the system should be balanced. The profit of the edge cloud is significantly influenced by the pricing policy when users care about the price a lot \cite{Pricing Zhao}. Literature \cite{Pricing} shed lights on the pricing and resource allocation problem in video caching systems from a game theoretic perspective.
\subsubsection{Scalability}
Scalability is an important feature that mobile edge networks provide compared to the legacy systems. The increasing number of mobile devices such as IoT devices will require scalability of services by applying load balancing mechanism \cite{MEC survey}. A cloud orchestrator which flexibly manage the edge node may provide network scalability \cite{virtual EPC}.

\subsubsection{Security}
The deployment of edge cloud servers is creating novel security challenges due to the exploiting of the mobile device information \cite{MEC Mtibaa}. The growing rate of the evolution of security solutions cannot keep up with the pace of the new security challenges. Many existing security protocols assume full connectivity \cite{Deng Security}, which is not realistic in mobile edge networks since many links are intermittent by default. A defense technique is proposed in \cite{MEC Mtibaa} for malicious node in mobile edge computing platforms called HoneyBot. The HoneyBot nodes can detect, track, and isolate D2D insider attacks. The speed and accuracy of this technique are impacted by the placement and number of HoneyBot nodes.

The security solutions for cloud computing may not suitable for edge computing because the working surroundings of edge devices will face many new threats different from well managed cloud \cite{Fog Security}. The authentication at different levels of gateways and smart meters is another important security issue. Some solutions have been proposed for the authentication problem, such as public key infrastructure (PKI) based solutions \cite{PKI}, Diffie-Hellman key exchange based solutions \cite{Diffie}.

Intrusion detection techniques are also required in fog computing. Some detection methods have been proposed for different applications \cite{Fog Security}. For example, signature-based method observes the behavior patterns and checks against a database of possible misbehaviors. The anomaly-based method detects intrusion by comparing an observed behavior with expected behavior to check the deviation.
\subsubsection{Privacy}
Privacy issues deal with hiding details. The content sharing and computation cooperation among users via D2D communication draw the concern on user's privacy. How to exploit the resources of mobile devices without invasion of privacy remains a challenge. In the typical use case of fog computing, the smart grid, the encryption of data from the smart meters and the aggregated result in Fog devices ensures data privacy. The original data can only be decrypted at the operation center \cite{Privacy grid}. In addition, many data privacy mechanisms have been developed for MCC to enforce privacy policies among collaborating mobile devices and to conceal the location of a set of clients \cite{MEC sec}. These mechanisms shed light on the design of privacy mechanisms for collaborative edge data centers in mobile edge networks. Most of the existing privacy solutions only require a trusted platform module (TPM), which can be deployed in edge data centers.

In the private networks cases such as personal cloudlets and corporate networks, the users and the edge data centers have a trustworthy relationship. Thus, privacy helper entities can be deployed in the edge data centers \cite{MEC sec}. Various data privacy mechanisms and services can be implemented on these helpers. In addition, the edge paradigm is helpful in strengthen the privacy feature of certain services. For instance, the anonymity of the users of location-based services can be protected by deploying a crowd-souring platform in a trusted edge server \cite{Privacy MCC}.
\subsubsection{User Participation}
A major idea in mobile edge networks is give the opportunity to utilize user terminals' available resources. Uses could be empowered with toolkits that enable users play an active role in technology design \cite{Cloud IoT}. However, this cooperation among users depends on users' willingness to participate. User incentive mechanisms need to be considered when designing computation offloading and caching strategies \cite{MEC Chen}.



\subsection{Discussions on Future Directions}

\subsubsection{Utilization of Wireless Big Data}
The wireless big data generated in the mobile edge networks is a valuable resources for analysis and design of the networks. Context aware approach also need the analysis of large amount of context information data. For example, user information big data can be utilized for popularity estimation in edge caching systems \cite{cache big data}. Thus, the full utilization of wireless big data will provide new opportunities in the performance of mobile networks.


\subsubsection{Online Caching}
The caching problem include two phases: cache placement and content delivery. Many works have been done on the cache updating during placement phase. However, more efficient caching update rules during the delivery phase, i.e. online caching, is a future direction of caching researches \cite{caching limits}.


\subsubsection{Context Awareness}
The mobile edge networks are advantageous in exploiting context information. The context provides information such as user location, other users in vicinity, and resources in the environment \cite{context Schilit}. The real time context aware applications could be accomplished by collaborations among MEC platforms \cite{MEC Nunna}. Different levels of context information (application, network and device level) could be used to proactively allocate resources \cite{Context Yang}.
\subsubsection{Smart User Association}
In cache enabled ultra dense HetNet, caching may change the way of user association other than the conventional nearest distance or SINR based method \cite{future direct}. The users may associate to the BS which caches the content it requested. In this way, the nearest BS may have strong interference on the users. Cache aware user association may overcome the backhaul capacity limitations and enhance users' QoE \cite{cache user association}. There could be more than one user access mode and users will select the best access mode for them \cite{user access fog}.


\subsubsection{Integration}
The architecture of mobile edge networks involve in various resources: computing, storage and communications. The efficient integration of these resources to achieve the optimal performance for all users and application is not concluded. More comprehensive resource allocation schemes need to be developed. Research on this topic may continue with the evolving of the network.

\section{Conclusion}
\label{sec:conc}
This paper surveys and summarizes the research efforts made on the mobile edge network, which is a paradigm integrating computing, caching and communication resources. The proposed architectures for edge computing and caching are presented including the ETSI MEC, Fog Computing, Cloudlet and edge caching. The related issues of computing, caching and communications are discussed respectively. For edge computing, the proposed schemes of computation offloading are extensively surveyed and classified. The issues of cooperations between the edge and the core as well as some existing edge computing platforms are presented. For edge caching, we make a detailed taxonomy on where, what and how to cache. Then the advances of communication techniques and the synergy with computing and caching are discussed. The novel applications and use cases are the driven force of the mobile edge network architecture. We summarize these applications and uses cases that mobile edge networks can fully enable. This new paradigm faces many challenges and opportunities. We point out the future research directions of this hot topic.

\section*{Acknowledgment}

The authors would like to thank the editor and the anonymous reviewers for their constructive comments and suggestions, which improve the quality of this paper.

\ifCLASSOPTIONcaptionsoff
  \newpage
\fi



\begin{thebibliography}{200}


\bibitem{Cisco}
    Cisco, ``Visual Networking Index," White paper, Feb. 2016 [Online]. Available: www.Cisco.com
\bibitem{MTC stat}
    H. Shariatmadari \emph{et al.}, ``Machine-type communications: current status and future perspectives toward 5G systems," \emph{IEEE Communications Magazine}, vol. 53, no. 9, pp. 10-17, Sep. 2015.
\bibitem{MTC arch}
    M. Condoluci, M. Dohler, G. Araniti, A. Molinaro and K. Zheng, ``Toward 5G densenets: architectural advances for effective machine-type communications over femtocells,"  \emph{IEEE Communications Magazine}, vol. 53, no. 1, pp. 134-141, Jan. 2015.
\bibitem{MTC que}
    F. M. Awuor and C. Y. Wang, ``Massive machine type communication in cellular system: A distributed queue approach," in \emph{2016 IEEE International Conference on Communications (ICC)}, Kuala Lumpur, 2016, pp. 1-7.
\bibitem{IoT in 5G}
    M. R. Palattella et al., ``Internet of Things in the 5G Era: Enablers, Architecture, and Business Models," \emph{IEEE Journal on Selected Areas in Communications}, vol. 34, no. 3, pp. 510-527, March 2016.
\bibitem{MC}
    M. Satyanarayanan, ``Mobile computing: the next decade," \emph{ACM SIGMOBILE Mobile Computing and Communications Review}, vol. 15, no. 2 pp. 2-10, Apr. 2011.
\bibitem{MCC app}
    A. u. R. Khan, M. Othman, S. A. Madani and S. U. Khan, ``A Survey of Mobile Cloud Computing Application Models," \emph{IEEE Communications Surveys \& Tutorials}, vol. 16, no. 1, pp. 393-413, First Quarter 2014.
\bibitem{MCC Zhang}
    W. Zhang, Y. Wen, K. Guan, D. Kilper, H. Luo and D. O. Wu, ``Energy-Optimal Mobile Cloud Computing under Stochastic Wireless Channel," \emph{IEEE Transactions on Wireless Communications}, vol. 12, no. 9, pp. 4569-4581, Sep. 2013.
\bibitem{MCC_Guan}
    L. Guan, X. Ke, M. Song and J. Song, ``A Survey of Research on Mobile Cloud Computing," in \emph{Computer and Information Science (ICIS), 2011 IEEE/ACIS 10th International Conference on}, Sanya, China, 2011, pp. 387-392.
\bibitem{MCC}
    H. T. Dinh, C. Lee, D. Niyato, and P. Wang, ``A survey of mobile cloud computing: architecture, applications, and approaches," \emph{Wireless communications and mobile computing}, vol. 13, no. 18, pp. 1587-1611, Dec. 2013.
\bibitem{MCC ZTE}
    X. Fan, J. Cao, and H. Mao, ``A survey of mobile cloud computing," \emph{ZTE Communications}, vol. 9, n. 1, pp. 4-8, 2011.
\bibitem{Hyrax}
    E. E. Marinelli, ``Hyrax: cloud computing on mobile devices using mapreduce," DTIC Document, Tech. Rep., 2009.
\bibitem{cloudlet}
    Y. Gao, W. Hu,  K. Ha, B. Amos, P. Pillai and M. Satyanarayanan, ``Are Cloudlets Necessary?" Technical Report CMU每CS每15每139, School of Computer Science, Carnegie Mellon University, Oct. 2015.
\bibitem{MCC access}
    A. Klein, C. Mannweiler, J. Schneider and H. D. Schotten, ``Access Schemes for Mobile Cloud Computing," in \emph{2010 Eleventh International Conference on Mobile Data Management}, Kansas City, MO, USA, 2010, pp. 387-392.
\bibitem{MCC Fern}
    N. Fernando, S. W. Loke, and  W. Rahayu, ``Mobile cloud computing: A survey," \emph{Future Generation Computer Systems}, vol. 29, no. 1, pp. 84-106, Jan., 2013.
\bibitem{MCC Auth}
    M. Alizadeh, S. Abolfazli, M. Zamani, S. Baharun and K. Sakurai, ``Authentication in mobile cloud computing: A survey," \emph{Journal of Network and Computer Applications}, vol. 61, pp. 59-80, Feb. 2016.
\bibitem{MEC ETSI}
    Y. C. Hu, M. Patel, D. Sabella,  N. Sprecher and V. Young, ``Mobile Edge Computing〞A Key Technology Towards 5G, " ETSI White Paper, 2015.
\bibitem{5G Vision}
    5G Infrastructure Public Private Partnership, ``5G Vision: The 5G Infrastructure Public Private Partnership: the next generation of communication networks and services," [Online]. Available: https://5g-ppp.eu/wp-content/uploads/2015/02/5G-Vision-Brochure-v1.pdf.
\bibitem{MEC survey}
    A. Ahmed and E. Ahmed,``A survey on mobile edge computing," in \emph{2016 10th International Conference on Intelligent Systems and Control (ISCO)}, Coimbatore, India, 2016, pp. 1-8.
\bibitem{MEC taxo}
    M. T. Beck, M. Werner, S. Feld and  S. Schimper, ``Mobile edge computing: A taxonomy," in \emph{Proc. of the Sixth International Conference on Advances in Future Internet}, Citeseer, 2014, pp. 48-54.
\bibitem{Fog role in IoT}
    F. Bonomi, R. Milito, J. Zhu, and S. Addepalli, ``Fog computing and its role in the internet of things," in \emph{Proceedings of the first edition of the MCC workshop on Mobile cloud computing}, 2012, pp. 13-16.
\bibitem{fog survey}
    S. Yi, C. Li, and Q. Li, ``A survey of fog computing: concepts, applications and issues," in \emph{Proceedings of the 2015 Workshop on Mobile Big Data, ACM.}, Hangzhou, China, 2015, pp. 37-42.
\bibitem{fog sec}	
    S. Yi, Z. Qin and Q. Li, ``Security and privacy issues of fog computing: A survey," in \emph{International Conference on Wireless Algorithms, Systems, and Applications}, Springer. 2015, pp. 685-695.

\bibitem{MEC sec}
    R. Roman, J. Lopez and M. Mambo, ``Mobile Edge Computing, Fog et al.: A Survey and Analysis of Security Threats and Challenges." \emph{arXiv preprint arXiv:1602.00484}, 2016.
\bibitem{vm based cloudlet}
    M. Satyanarayanan, P. Bahl, R. Caceres and N. Davies, ``The Case for VM-Based Cloudlets in Mobile Computing,"  \emph{IEEE Pervasive Computing}, vol. 8, no. 4, pp. 14-23, Oct.-Dec. 2009.
\bibitem{cache in air}
    X. Wang, M. Chen, T. Taleb, A. Ksentini and V. C. M. Leung, ``Cache in the air: exploiting content caching and delivery techniques for 5G systems," \emph{IEEE Communications Magazine}, vol. 52, no. 2, pp. 131-139, February 2014.

\bibitem{web caching}
    W. Ali, S. M. Shamsuddin and  A. S. Ismail, ``A survey of Web caching and prefetching," \emph{Int. J. Advance. Soft Comput. Appl}, vol. 3, no. 1, pp. 18-44, Mar. 2011.
\bibitem{web cache replace}
    S. Podlipnig and L. B\"osz\"ormenyi, ``A survey of web cache replacement strategies," \emph{ACM Computing Surveys (CSUR)}, vol. 35, no. 4, pp. 374-398, Dec. 2003.
\bibitem{Cache ICN}
    M. Zhang, H. Luo and H. Zhang, ``A Survey of Caching Mechanisms in Information-Centric Networking," \emph{IEEE Communications Surveys \& Tutorials}, vol. 17, no. 3, pp. 1473-1499, Third Quarter 2015.
\bibitem{Cache ICN EE}
    C. Fang, F. R. Yu, T. Huang, J. Liu and Y. Liu, ``A survey of energy-efficient caching in information-centric networking," \emph{IEEE Communications Magazine}, vol. 52, no. 11, pp. 122-129, Nov. 2014.
\bibitem{NFV vEPC}
    H. Hawilo, A. Shami, M. Mirahmadi and R. Asal, ``NFV: state of the art, challenges, and implementation in next generation mobile networks (vEPC)," \emph{IEEE Network}, vol. 28, no. 6, pp. 18-26, Nov.-Dec. 2014.

\bibitem{DOC}
    X. Zhang et al., ``Macro-assisted data-only carrier for 5G green cellular systems," \emph{IEEE Communications Magazine}, vol. 53, no. 5, pp. 223-231, May 2015.
\bibitem{SDMN}
    K. Pentikousis, Y. Wang and W. Hu, ``Mobileflow: Toward software-defined mobile networks," \emph{IEEE Communications Magazine}, vol. 51, no. 7, pp. 44-53, Jul. 2013.

\bibitem{5G survey}
    M. Agiwal, A. Roy and N. Saxena, ``Next Generation 5G Wireless Networks: A Comprehensive Survey," \emph{IEEE Communications Surveys \& Tutorials}, vol. 18, no. 3, pp. 1617-1655, Third Quarter 2016.
\bibitem{5G dir}
    F. Boccardi, R. W. Heath, A. Lozano, T. L. Marzetta and P. Popovski, ``Five disruptive technology directions for 5G," \emph{IEEE Communications Magazine}, vol. 52, no. 2, pp. 74-80, Feb. 2014.

\bibitem{Fog mec cloudlet}
    G. I. Klas, ``Fog computing and mobile edge cloud gain momentum open fog consortium etsi mec and cloudlets", 2015, [online] Available: http://yucianga.info/?p=938.
\bibitem{fog iot}
     F. Bonomi et al, ``Fog computing: A platform for internet of things and analytics," in \emph{Big Data and Internet of Things: A Roadmap for Smart Environments}, Springer International Publishing, 2014, pp. 169-186.
\bibitem{ST model}
    S. Wang, X. Zhang, J. Zhang, J. Feng, W. Wang and K. Xin, ``An Approach for Spatial-Temporal Traffic Modeling in Mobile Cellular Networks," in \emph{2015 27th International Teletraffic Congress}, Ghent, 2015, pp. 203-209.
\bibitem{living on edge}
    E. Ba\c{s}tu\u{g}, M. Bennis and M. Debbah, ``Living on the edge: The role of proactive caching in 5G wireless networks," \emph{IEEE Communications Magazine}, vol. 52, no. 8, pp. 82-89, Aug. 2014.
\bibitem{MEC white paper}
    European Telecommunications Standards Institute, ``Mobile-Edge Computing Introductory Technical White Paper," White Paper, 2015.

\bibitem{Fog network}
    M. Chiang, ``Fog networking: An overview on research opportunities," \emph{arXiv preprint arXiv:1601.00835}, 2016.
\bibitem{Fog IoT Overview}
    M. Chiang; T. Zhang, ``Fog and IoT: An Overview of Research Opportunities," \emph{IEEE Internet of Things Journal}, to appear.
\bibitem{Fog RAN}
    R. Tandon and O. Simeone, "Harnessing cloud and edge synergies: toward an information theory of fog radio access networks," in IEEE Communications Magazine, vol. 54, no. 8, pp. 44-50, August 2016.
\bibitem{cloudlet eco}
    M. Satyanarayanan et al., ``An open ecosystem for mobile-cloud convergence," \emph{IEEE Communications Magazine}, vol. 53, no. 3, pp. 63-70, Mar. 2015.
\bibitem{MEC cloudlet}
    Y. Jararweh, A. Doulat, O. AlQudah, E. Ahmed, M. Al-Ayyoub and E. Benkhelifa, ``The future of mobile cloud computing: Integrating cloudlets and Mobile Edge Computing," in \emph{2016 23rd International Conference on Telecommunications (ICT)}, Thessaloniki, 2016, pp. 1-5.
\bibitem{MEC Rimal}
    B. P. Rimal, D. Pham Van and M. Maier, ``Mobile-edge computing vs. centralized cloud computing in fiber-wireless access networks," in \emph{2016 IEEE Conference on Computer Communications Workshops (INFOCOM WKSHPS)}, San Francisco, CA, 2016, pp. 991-996.
\bibitem{mec quantify}
    W. Hu et al., ``Quantifying the Impact of Edge Computing on Mobile Applications," in \emph{Proceedings of the 7th ACM SIGOPS Asia-Pacific Workshop on Systems}, Hong Kong, 2016, pp. 1-8.
\bibitem{MEN benefit}
    M. Amardeep et al. ``How beneficial are intermediate layer Data Centers in Mobile Edge Networks?" in |emph{Workshops on Fog and Mobile Edge Computing (FMEC 2016)}, 2016, pp. 222-229.
\bibitem{Fog Energy}
    F. Jalali, K. Hinton, R. Ayre, T. Alpcan and R. S. Tucker, ``Fog Computing May Help to Save Energy in Cloud Computing," \emph{IEEE Journal on Selected Areas in Communications}, vol. 34, no. 5, pp. 1728-1739, May 2016.
\bibitem{MEC Nunna}
    S. Nunna et al., ``Enabling Real-Time Context-Aware Collaboration through 5G and Mobile Edge Computing," in \emph{Information Technology - New Generations (ITNG), 2015 12th International Conference on}, Las Vegas, NV, 2015, pp. 601-605.
\bibitem{MEC in 5G}
    K. Zhang et al., ``Energy-Efficient Offloading for Mobile Edge Computing in 5G Heterogeneous Networks," \emph{IEEE Access}, vol. 4, pp. 5896-5907, 2016.
\bibitem{ennergy Fog}
    P. Borylo, A. Lason, J. Rzasa, A. Szymanski and A. Jajszczyk, ``Energy-aware fog and cloud interplay supported by wide area software defined networking," in \emph{2016 IEEE International Conference on Communications (ICC)}, Kuala Lumpur, 2016, pp. 1-7.
\bibitem{metis 5G}
    A. Osseiran et al., ``Scenarios for 5G mobile and wireless communications: the vision of the METIS project," \emph{IEEE Communications Magazine}, vol. 52, no. 5, pp. 26-35, May 2014.
\bibitem{5G considerations}
    P. K. Agyapong, M. Iwamura, D. Staehle, W. Kiess and A. Benjebbour, ``Design considerations for a 5G network architecture," \emph{IEEE Communications Magazine}, vol. 52, no. 11, pp. 65-75, Nov. 2014.
\bibitem{Fog Cloud Souza}
    V. B. C. Souza, W. Ram赤rez, X. Masip-Bruin, E. Mar赤n-Tordera, G. Ren and G. Tashakor, ``Handling service allocation in combined Fog-cloud scenarios," in \emph{2016 IEEE International Conference on Communications (ICC)}, Kuala Lumpur, 2016, pp. 1-5.
\bibitem{MEC power control}
    P. Mach and Z. Becvar, ``Cloud-aware power control for real-time application offloading in mobile edge computing," \emph{Transactions on Emerging Telecommunications Technologies}, vol. 27, no. 5, pp. 648-661, Dec. 2015.
\bibitem{MEC delay opt}
    J. Liu, Y. Mao, J. Zhang and K. B. Letaief, ``Delay-optimal computation task scheduling for mobile-edge computing systems," in \emph{2016 IEEE International Symposium on Information Theory (ISIT)}, Barcelona, 2016, pp. 1451-1455.
\bibitem{MEC min delay}
    J. O. Fajardo, I. Taboada and F. Liberal, ``Radio-aware service-level scheduling to minimize downlink traffic delay through Mobile Edge Computing," In \emph{International Conference on Mobile Networks and Management}, 2015, pp. 121-134.
\bibitem{MEC Mao}
    Y. Mao; J. Zhang; K. B. Letaief, ``Dynamic Computation Offloading for Mobile-Edge Computing with Energy Harvesting Devices," \emph{IEEE Journal on Selected Areas in Communications}, vol.PP, no.99, pp.1-1.
\bibitem{multiuser clustering}
    J. Oueis, E. C. Strinati and S. Barbarossa, ``Distributed mobile cloud computing: A multi-user clustering solution," in \emph{2016 IEEE International Conference on Communications (ICC)}, Kuala Lumpur, 2016, pp. 1-6.
\bibitem{MEC Zhang}
    K. Zhang, Y. Mao, S. Leng, A. Vinel and Y. Zhang, "Delay constrained offloading for Mobile Edge Computing in cloud-enabled vehicular networks," 2016 8th International Workshop on Resilient Networks Design and Modeling (RNDM), Halmstad, Sweden, 2016, pp. 288-294.
\bibitem{femtoclouds}
     K. Habak, M. Ammar, K. A. Harras and E. Zegura, ``Femto Clouds: Leveraging Mobile Devices to Provide Cloud Service at the Edge," in \emph{2015 IEEE 8th International Conference on Cloud Computing}, New York City, NY, 2015, pp. 9-16.
\bibitem{MEC Partial offloading}
    Y. Wang, M. Sheng, X. Wang, L. Wang and J. Li, ``Mobile-Edge Computing: Partial Computation Offloading Using Dynamic Voltage Scaling," \emph{IEEE Transactions on Communications}, vol. 64, no. 10, pp. 4268-4282, Oct. 2016.
\bibitem{MEC Chen}
    X. Chen, L. Jiao, W. Li and X. Fu, "Efficient Multi-User Computation Offloading for Mobile-Edge Cloud Computing," \emph{IEEE/ACM Transactions on Networking}, vol. 24, no. 5, pp. 2795-2808, Oct. 2016.	
\bibitem{MEC Sard}
    S. Sardellitti, G. Scutari and S. Barbarossa, ``Joint Optimization of Radio and Computational Resources for Multicell Mobile-Edge Computing," \emph{IEEE Transactions on Signal and Information Processing over Networks}, vol. 1, no. 2, pp. 89-103, Jun. 2015.

\bibitem{MEC Huang}
    C. You, K. Huang, H. Chae and B. H. Kim, ``Energy-Efficient Resource Allocation for Mobile-Edge Computation Offloading," \emph{arXiv preprint arXiv:1605.08518}, 2016.

\bibitem{MEC Kety}
    I. Ketyk\'o, L. Kecsk\'es, C. Nemes and L. Farkas, ``Multi-user computation offloading as Multiple Knapsack Problem for 5G Mobile Edge Computing," in \emph{2016 European Conference on Networks and Communications (EuCNC)}, Athens, 2016, pp. 225-229.
\bibitem{MEC  sequen}
    M. Deng, H. Tian and X. Lyu, ``Adaptive sequential offloading game for multi-cell Mobile Edge Computing," in \emph{2016 23rd International Conference on Telecommunications (ICT)}, Thessaloniki, 2016, pp. 1-5.

\bibitem{MEC Opportunistic}
    M. H. u. Rehman, C. Sun, T. Y. Wah, A. Iqbal and P. P. Jayaraman, ``Opportunistic Computation Offloading in Mobile Edge Cloud Computing Environments," in \emph{2016 17th IEEE International Conference on Mobile Data Management (MDM)}, Porto, 2016, pp. 208-213.
\bibitem{REPLISOM}
    S. Abdelwahab, B. Hamdaoui, M. Guizani and T. Znati, ``Replisom: Disciplined Tiny Memory Replication for Massive IoT Devices in LTE Edge Cloud," \emph{IEEE Internet of Things Journal}, vol. 3, no. 3, pp. 327-338, June 2016.
\bibitem{ME VoLTE}
    M. T. Beck, S. Feld, A. Fichtner, C. Linnhoff-Popien and T. Schimper, ``ME-VoLTE: Network functions for energy-efficient video transcoding at the mobile edge," in \emph{Intelligence in Next Generation Networks (ICIN), 2015 18th International Conference on}, Paris, 2015, pp. 38-44.
\bibitem{Load distribution}
    J. Oueis, E. C. Strinati and S. Barbarossa, ``The Fog Balancing: Load Distribution for Small Cell Cloud Computing," in \emph{2015 IEEE 81st Vehicular Technology Conference (VTC Spring)}, Glasgow, 2015, pp. 1-6.
\bibitem{MEC task offloading}
    L. Tianze, W. Muqing and Z. Min, ``Consumption considered optimal scheme for task offloading in mobile edge computing," in \emph{2016 23rd International Conference on Telecommunications (ICT)}, Thessaloniki, 2016, pp. 1-6.
\bibitem{Mobility Chen}
     M. Chen, Y. Hao,  M. Qiu, J. Song, D. Wu and I. Humar, ``Mobility-Aware Caching and Computation Offloading in 5G Ultra-Dense Cellular Networks," \emph{Sensors}, vol. 16, no. 7, pp. 974-986, Jun. 2016.

\bibitem{mobility model}
    Y. Li and W. Wang, ``Can mobile cloudlets support mobile applications?" in \emph{IEEE INFOCOM 2014 - IEEE Conference on Computer Communications}, Toronto, ON, 2014, pp. 1060-1068.
\bibitem{Fog cloud}
    R. Deng, R. Lu, C. Lai and T. H. Luan, ``Towards power consumption-delay tradeoff by workload allocation in cloud-fog computing," in \emph{2015 IEEE International Conference on Communications (ICC)}, London, 2015, pp. 3909-3914.
\bibitem{edge assisted cloud}
    I. Farris, L. Militano, M. Nitti, L. Atzori and A. Iera, ``Federated edge-assisted mobile clouds for service provisioning in heterogeneous IoT environments," in \emph{Internet of Things (WF-IoT), 2015 IEEE 2nd World Forum on}, Milan, 2015, pp. 591-596.
\bibitem{RACS}	
    Intel and Nokia Siemens Networks, ``Increasing mobile operators＊ value proposition with edge computing," [Online]. Available:  http://www.intel.co.id/content/dam/www/public/us/en/documents/
    technology-briefs/edge-computing-tech-brief.pdf.
\bibitem{Fog Industrial IoT}
    V. Gazis, A. Leonardi, K. Mathioudakis, K. Sasloglou, P. Kikiras and R. Sudhaakar, ``Components of fog computing in an industrial internet of things context," in \emph{Sensing, Communication, and Networking - Workshops (SECON Workshops), 2015 12th Annual IEEE International Conference on}, Seattle, WA, 2015, pp. 1-6.
\bibitem{Fog platorm D2D}
    C. Vallati, A. Virdis, E. Mingozzi and G. Stea, ``Exploiting LTE D2D communications in M2M Fog platforms: Deployment and practical issues," in \emph{Internet of Things (WF-IoT), 2015 IEEE 2nd World Forum on}, Milan, 2015, pp. 585-590.
\bibitem{cache dynamic traffic}
    B. Xia; C. Yang; T. Cao, ``Modeling and Analysis for Cache-enabled Networks with Dynamic Traffic," \emph{IEEE Communications Letters},to appear.

\bibitem{video aware scheduling} 	
    H. Ahlehagh and S. Dey, ``Video-Aware Scheduling and Caching in the Radio Access Network," \emph{IEEE/ACM Transactions on Networking}, vol. 22, no. 5, pp. 1444-1462, Oct. 2014.


\bibitem{Proactive Storage}
    J. Gu, W. Wang, A. Huang and H. Shan, ``Proactive storage at caching-enable base stations in cellular networks," in \emph{2013 IEEE 24th Annual International Symposium on Personal, Indoor, and Mobile Radio Communications (PIMRC)}, London, 2013, pp. 1543-1547.
\bibitem{caching modeling bustug}
    E. Ba\c{s}tu\u{g}, M. Bennis and M. Debbah, ``Cache-enabled small cell networks: Modeling and tradeoffs," in \emph{2014 11th International Symposium on Wireless Communications Systems (ISWCS)}, Barcelona, 2014, pp. 649-653.
\bibitem{learning based caching}
    P. Blasco and D. G\"und\"uz, ``Learning-based optimization of cache content in a small cell base station," in \emph{2014 IEEE International Conference on Communications (ICC)}, Sydney, NSW, 2014, pp. 1897-1903.
\bibitem{femtocaching}
    N. Golrezaei, K. Shanmugam, A. G. Dimakis, A. F. Molisch and G. Caire, ``FemtoCaching: Wireless video content delivery through distributed caching helpers," in \emph{2012 Proceedings IEEE INFOCOM, Orlando}, FL, 2012, pp. 1107-1115.
\bibitem{multicast cache}
    K. Poularakis, G. Iosifidis, V. Sourlas and L. Tassiulas, ``Multicast-aware caching for small cell networks," in \emph{2014 IEEE Wireless Communications and Networking Conference (WCNC)}, Istanbul, 2014, pp. 2300-2305.
\bibitem{caching and routing}
    K. Poularakis, G. Iosifidis and L. Tassiulas, ``Approximation caching and routing algorithms for massive mobile data delivery," in \emph{2013 IEEE Global Communications Conference (GLOBECOM)}, Atlanta, GA, 2013, pp. 3534-3539.
\bibitem{D2D caching social}
    B. Bai, L. Wang, Z. Han, W. Chen and T. Svensson, ``Caching based socially-aware D2D communications in wireless content delivery networks: a hypergraph framework," \emph{IEEE Wireless Communications}, vol. 23, no. 4, pp. 74-81, Aug. 2016.
\bibitem{D2D caching Yang}
    B. Chen, C. Yang and G. Wang, ``Cooperative Device-to-Device Communications with Caching," in \emph{2016 IEEE 83rd Vehicular Technology Conference (VTC Spring)}, Nanjing, 2016, pp. 1-5.

\bibitem{cache misconceptions}
    G. Paschos, E. Ba\c{s}tu\u{g}, I. Land, G. Caire and M. Debbah, ``Wireless caching: technical misconceptions and business barriers," \emph{IEEE Communications Magazine}, vol. 54, no. 8, pp. 16-22, Aug. 2016.
\bibitem{Zipf}
    L. Breslau, Pei Cao, Li Fan, G. Phillips and S. Shenker, ``Web caching and Zipf-like distributions: evidence and implications," in \emph{Proceedings of INFOCOM '99. Eighteenth Annual Joint Conference of the IEEE Computer and Communications Societies}, New York, NY, 1999, pp. 126-134.
\bibitem{SNM}
    S. Traverso et al. ``Temporal locality in today's content caching: why it matters and how to model it," \emph{ACM SIGCOMM}, vol. 43, no. 5, pp. 5-12, Oct. 2013.
\bibitem{video popularity}
    M. Cha, H. Kwak, P. Rodriguez, Y. Y. Ahn and S. Moon, ``Analyzing the Video Popularity Characteristics of Large-Scale User Generated Content Systems," \emph{IEEE/ACM Transactions on Networking}, vol. 17, no. 5, pp. 1357-1370, Oct. 2009.
\bibitem{Caching Policy survey}
    A. Ioannou and S. Weber, ``A Survey of Caching Policies and Forwarding Mechanisms in Information-Centric Networking," \emph{IEEE Communications Surveys \& Tutorials}, vol. 18, no. 4, pp. 2847-2886, Fourthquarter 2016.
\bibitem{LRU}
    N. Laoutaris, ``A closed-form method for LRU replacement under generalized power-law demand," \emph{arXiv preprint arXiv:0705.1970}, 2007.

\bibitem{Learning caching}
    A. Sengupta, S. Amuru, R. Tandon, R. M. Buehrer and T. C. Clancy, ``Learning distributed caching strategies in small cell networks," in \emph{2014 11th International Symposium on Wireless Communications Systems (ISWCS)}, Barcelona, 2014, pp. 917-921.
\bibitem{cache replace}
    J. Gu, W. Wang, A. Huang, H. Shan and Z. Zhang, ``Distributed cache replacement for caching-enable base stations in cellular networks," in \emph{2014 IEEE International Conference on Communications (ICC)}, Sydney, NSW, 2014, pp. 2648-2653.
\bibitem{Caching Distri}
    S. Borst, V. Gupta and A. Walid, ``Distributed Caching Algorithms for Content Distribution Networks," in \emph{2010 Proceedings IEEE INFOCOM}, San Diego, CA, 2010, pp. 1-9.

\bibitem{cooperative caching and deli}
    W. Jiang; G. Feng; S. Qin, ``Optimal Cooperative Content Caching and Delivery Policy for Heterogeneous Cellular Networks," \emph{IEEE Transactions on Mobile Computing}, to appear, 2016.
\bibitem{cache SVC}
    R. Yu et al., ``Enhancing software-defined RAN with collaborative caching and scalable video coding," in \emph{2016 IEEE International Conference on Communications (ICC)}, Kuala Lumpur, 2016, pp. 1-6.
\bibitem{cache redundancy}
    S. Wang, X. Zhang, K. Yang, L. Wang and W. Wang, "Distributed edge caching scheme considering the tradeoff between the diversity and redundancy of cached content," in \emph{2015 IEEE/CIC International Conference on Communications in China (ICCC)}, Shenzhen, 2015, pp. 1-5.
\bibitem{network coding}
    P. Ostovari, A. Khreishah and J. Wu, ``Cache content placement using triangular network coding," in \emph{2013 IEEE Wireless Communications and Networking Conference (WCNC)}, Shanghai, 2013, pp. 1375-1380.
\bibitem{IoT Caching}
    S. Vural, P. Navaratnam, N. Wang, C. Wang, L. Dong and R. Tafazolli, ``In-network caching of Internet-of-Things data," in \emph{2014 IEEE International Conference on Communications (ICC)}, Sydney, NSW, 2014, pp. 3185-3190.
\bibitem{mobility method}
    R. Wang, X. Peng, J. Zhang and K. B. Letaief, ``Mobility-aware caching for content-centric wireless networks: modeling and methodology," \emph{IEEE Communications Magazine}, vol. 54, no. 8, pp. 77-83, Aug. 2016.
\bibitem{MobiCacher}
    Y. Guan, Y. Xiao, H. Feng, C. C. Shen and L. J. Cimini, ``MobiCacher: Mobility-aware content caching in small-cell networks," in \emph{2014 IEEE Global Communications Conference}, Austin, TX, 2014, pp. 4537-4542.
\bibitem{Mobility on hit}
    C. Jarray and A. Giovanidis, ``The effects of mobility on the hit performance of cached D2D networks," in \emph{2016 14th International Symposium on Modeling and Optimization in Mobile, Ad Hoc, and Wireless Networks (WiOpt)}, Tempe, AZ, 2016, pp. 1-8.
\bibitem{exploit mobility}
    K. Poularakis and L. Tassiulas, ``Exploiting user mobility for wireless content delivery," in \emph{2013 IEEE International Symposium on Information Theory}, Istanbul, 2013, pp. 1017-1021.
\bibitem{mobility prediction}
    H. Li and D. Hu, ``Mobility prediction based seamless RAN-cache handover in HetNet," in \emph{2016 IEEE Wireless Communications and Networking Conference}, Doha, 2016, pp. 1-7.
\bibitem{Edge caching mobility predic}
    A. S. Gomes, B. Sousa, D. Palma, V. Fonseca, Z. Zhao, E. Monteiro and L. Cordeiro, ``Edge caching with mobility prediction in virtualized LTE mobile networks" \emph{Future Generation Computer Systems}, to appear, 2016.
\bibitem{caching latency tradeoff}
    R. Tandon and O. Simeone, ``Cloud-aided wireless networks with edge caching: Fundamental latency trade-offs in fog Radio Access Networks," in \emph{2016 IEEE International Symposium on Information Theory (ISIT)}, Barcelona, 2016, pp. 2029-2033.
\bibitem{Hetnet caching}
    D. Liu and C. Yang, "Cache-enabled heterogeneous cellular networks: Comparison and tradeoffs," in \emph{2016 IEEE International Conference on Communications (ICC)}, Kuala Lumpur, 2016, pp. 1-6.
\bibitem{caching energy Yang}
    D. Liu and C. Yang, ``Will caching at base station improve energy efficiency of downlink transmission?," in \emph{2014 IEEE Global Conference on Signal and Information Processing (GlobalSIP)}, Atlanta, GA, 2014, pp. 173-177.
\bibitem{cache SD Hetnet}
    J. Zhang, X. Zhang and W. Wang, ``Cache-Enabled Software Defined Heterogeneous Networks for Green and Flexible 5G Networks," \emph{IEEE Access}, vol. 4, pp. 3591-3604, 2016.
\bibitem{MEC 5G}
    J. Q. Fajardo et al., ``Introducing mobile edge computing capabilities through distributed 5G cloud enabled small cells," \emph{Mobile networks and applications},  vol. 21, no. 4, pp. 564-574, 2016.
\bibitem{caching mmW}
    J. Qiao, Y. He and X. S. Shen, ``Proactive Caching for Mobile Video Streaming in Millimeter Wave 5G Networks," \emph{IEEE Transactions on Wireless Communications}, vol. 15, no. 10, pp. 7187-7198, Oct. 2016.
\bibitem{D2D LTE}
    K. Doppler, M. Rinne, C. Wijting, C. B. Ribeiro and K. Hugl, ``Device-to-device communication as an underlay to LTE-advanced networks," \emph{IEEE Communications Magazine}, vol. 47, no. 12, pp. 42-49, Dec. 2009.
\bibitem{D2D design aspects}
    G. Fodor et al., ``Design aspects of network assisted device-to-device communications," \emph{IEEE Communications Magazine}, vol. 50, no. 3, pp. 170-177, Mar. 2012.
\bibitem{D2D offloading}
    W. Cao; G. Feng; S. Qin; M. Yan, ``Cellular Offloading in Heterogeneous Mobile Networks with D2D Communication Assistance," \emph{IEEE Transactions on Vehicular Technology}, vol.PP, no.99, pp.1-1.
\bibitem{D2D 3GPP}
    3GPP TR 22.803, ``Feasibility Study for Proximity Services (ProSe)," (Release 12), Jun. 2013.
\bibitem{D2D SC}
    A. Laya, K. Wang, A. A. Widaa, J. Alonso-Zarate, J. Markendahl and L. Alonso, ``Device-to-device communications and small cells: enabling spectrum reuse for dense networks," \emph{IEEE Wireless Communications}, vol. 21, no. 4, pp. 98-105, Aug. 2014.
\bibitem{IA}
    N. Zhao, X. Liu, F. R. Yu, M. Li and V. C. M. Leung, ``Communications, caching, and computing oriented small cell networks with interference alignment," \emph{IEEE Communications Magazine}, vol. 54, no. 9, pp. 29-35, Sep. 2016.
\bibitem{VRE}
    C. Liu et al., ``Video content redundancy elimination based on the convergence of computing, communication and cache," in \emph{2016 IEEE Global Communications Conference (GLOBECOM)}, to appear.
\bibitem{synergy}
    S. Andreev et al., ``Exploring synergy between communications, caching, and computing in 5G-grade deployments," \emph{IEEE Communications Magazine}, vol. 54, no. 8, pp. 60-69, Aug. 2016.

\bibitem{web Zhu}
    J. Zhu, D. S. Chan, M. S. Prabhu, P. Natarajan, H. Hu, and F. Bonomi, ``Improving web sites performance using edge servers in fog computing architecture,§ in \emph{Service Oriented System Engineering (SOSE), 2013 IEEE 7th International Symposium on}, Redwood City, 2013, pp. 320每323.
\bibitem{MEC scenarios}
    ETSI (2015), ``Mobile-Edge Computing (MEC); Service Scenarios," [Online]. Available: http://www.etsi.org/deliver/etsi\_gs/MEC-IEG/
    001\_099/004/01.01.01\_60/gs\_MEC-IEG004v010101p.pdf.
\bibitem{streaming}
    O. M\"akinen, ``Streaming at the Edge: Local Service Concepts Utilizing Mobile Edge Computing," in \emph{Next Generation Mobile Applications, Services and Technologies, 2015 9th International Conference on}, Cambridge, 2015, pp. 1-6.
\bibitem{IoT survey}   	
    A. Al-Fuqaha, M. Guizani, M. Mohammadi, M. Aledhari and M. Ayyash, ``Internet of Things: A Survey on Enabling Technologies, Protocols, and Applications," \emph{IEEE Communications Surveys \& Tutorials}, vol. 17, no. 4, pp. 2347-2376, Fourth Quarter 2015.
\bibitem{Edge Analytics}
    M. Satyanarayanan et al., ``Edge Analytics in the Internet of Things," \emph{IEEE Pervasive Computing}, vol. 14, no. 2, pp. 24-31, Apr.-June 2015.
\bibitem{MEC Sabella}
    D. Sabella, A. Vaillant, P. Kuure, U. Rauschenbach and F. Giust, "Mobile-Edge Computing Architecture: The role of MEC in the Internet of Things," in IEEE Consumer Electronics Magazine, vol. 5, no. 4, pp. 84-91, Oct. 2016.
\bibitem{MEC Corcoran}
    P. Corcoran and S. K. Datta, ``Mobile-Edge Computing and the Internet of Things for Consumers: Extending cloud computing and services to the edge of the network," \emph{IEEE Consumer Electronics Magazine}, vol. 5, no. 4, pp. 73-74, Oct. 2016.

\bibitem{Fog IoT Potential}
    A. V. Dastjerdi and R. Buyya, ``Fog Computing: Helping the Internet of Things Realize Its Potential," \emph{Computer}, vol. 49, no. 8, pp. 112-116, Aug. 2016.
\bibitem{MEC smart grid}
    N. Kumar, S. Zeadally and J. J. P. C. Rodrigues, ``Vehicular delay-tolerant networks for smart grid data management using mobile edge computing," \emph{IEEE Communications Magazine}, vol. 54, no. 10, pp. 60-66, Oct. 2016.
\bibitem{MEC smart home}
    C. Vallati, A. Virdis, E. Mingozzi and G. Stea, ``Mobile-Edge Computing Come Home Connecting things in future smart homes using LTE device-to-device communications," \emph{IEEE Consumer Electronics Magazine}, vol. 5, no. 4, pp. 77-83, Oct. 2016.
\bibitem{MEC smart city}
    M. Sapienza, E. Guardo, M. Cavallo, G. La Torre, G. Leombruno and O. Tomarchio, ``Solving Critical Events through Mobile Edge Computing: An Approach for Smart Cities," in \emph{2016 IEEE International Conference on Smart Computing (SMARTCOMP)}, St. Louis, MO, 2016, pp. 1-5.
\bibitem{cloudlet cognitive}
    M. Satyanarayanan, Z. Chen, K. Ha, W. Hu, W. Richter and P. Pillai, ``Cloudlets: at the leading edge of mobile-cloud convergence," in \emph{Mobile Computing, Applications and Services (MobiCASE), 2014 6th International Conference on}, Austin, TX, 2014, pp. 1-9.
\bibitem{Cloud IoT}
    A. Botta, W. de Donato, V. Persico and A. Pescap谷, ``On the Integration of Cloud Computing and Internet of Things," in \emph{Future Internet of Things and Cloud (FiCloud), 2014 International Conference on}, Barcelona, 2014, pp. 23-30.
\bibitem{AT&T}
    AT\&T, ``Domain 2.0 Vision White Paper," \emph{white paper}, Aug. 2013.
\bibitem{MEC SDMEC}
    Y. Jararweh, A. Doulat, A. Darabseh, M. Alsmirat, M. Al-Ayyoub and E. Benkhelifa, ``SDMEC: Software Defined System for Mobile Edge Computing," in \emph{2016 IEEE International Conference on Cloud Engineering Workshop (IC2EW)}, Berlin, 2016, pp. 88-93.
\bibitem{MEC Salman}
    O. Salman, I. Elhajj, A. Kayssi and A. Chehab, ``Edge computing enabling the Internet of Things," in \emph{Internet of Things (WF-IoT), 2015 IEEE 2nd World Forum on}, Milan, 2015, pp. 603-608.

\bibitem{VNE}	
    M. T. Beck and M. Marco, ``Mobile Edge Computing: Challenges for Future Virtual Network Embedding Algorithms." in \emph{Proceedings of the 8th International Conference on Advanced Engineering Computing and Applications in Sciences (ADVCOMP)}, 2014, pp. 65-70.
\bibitem{ETSI NFV}
    ETSI, ISGNFV. ``Network Functions Virtualisation每Network Operator Perspectives on Industry Progress." Updated White Paper, 2013.


\bibitem{Pricing Zhao}
    T. Zhao, S. Zhou, X. Guo, Y. Zhao and Z. Niu, ``Pricing policy and computational resource provisioning for delay-aware mobile edge computing," in \emph{2016 IEEE/CIC International Conference on Communications in China (ICCC)}, Chengdu, 2016, pp. 1-6.
\bibitem{Pricing}
    J. Li, H. Chen, Y. Chen, Z. Lin, B. Vucetic and L. Hanzo, ``Pricing and Resource Allocation via Game Theory for a Small-Cell Video Caching System," \emph{IEEE Journal on Selected Areas in Communications}, vol. 34, no. 8, pp. 2115-2129, Aug. 2016.
\bibitem{virtual EPC}
     E. Cau et al., ``Efficient Exploitation of Mobile Edge Computing for Virtualized 5G in EPC Architectures," in \emph{2016 4th IEEE International Conference on Mobile Cloud Computing, Services, and Engineering (MobileCloud)}, Oxford, 2016, pp. 100-109.

\bibitem{MEC Mtibaa}
    A. Mtibaa, K. Harras and H. Alnuweiri, ``Friend or Foe? Detecting and Isolating Malicious Nodes in Mobile Edge Computing Platforms," in \emph{2015 IEEE 7th International Conference on Cloud Computing Technology and Science (CloudCom)}, Vancouver, BC, 2015, pp. 42-49.
\bibitem{Deng Security}
    Hongmei Deng, Wei Li and D. P. Agrawal, ``Routing security in wireless ad hoc networks," \emph{IEEE Communications Magazine}, vol. 40, no. 10, pp. 70-75, Oct. 2002.
\bibitem{Fog Security}
    I. Stojmenovic and S. Wen, ``The Fog computing paradigm: Scenarios and security issues," in \emph{Computer Science and Information Systems (FedCSIS), 2014 Federated Conference on}, Warsaw, 2014, pp. 1-8.
\bibitem{PKI}
    Y. W. Law, M. Palaniswami, G. Kounga and A. Lo, "WAKE: Key management scheme for wide-area measurement systems in smart grid," in IEEE Communications Magazine, vol. 51, no. 1, pp. 34-41, Jan. 2013.
\bibitem{Diffie}
    Z. M. Fadlullah, M. M. Fouda, N. Kato, A. Takeuchi, N. Iwasaki and Y. Nozaki, ``Toward intelligent machine-to-machine communications in smart grid," \emph{IEEE Communications Magazine}, vol. 49, no. 4, pp. 60-65, Apr. 2011.
\bibitem{Privacy grid}
    R. Lu, X. Liang, X. Li, X. Lin and X. Shen, ``EPPA: An Efficient and Privacy-Preserving Aggregation Scheme for Secure Smart Grid Communications," \emph{IEEE Transactions on Parallel and Distributed Systems}, vol. 23, no. 9, pp. 1621-1631, Sep. 2012.
\bibitem{Privacy MCC}
    J. Abdo, J. Demerjian, H. Chaouchi, T. Atechian and
    C. Bassil, ``Privacy using Mobile Cloud Computing," in \emph{Proceedings of the 5th International Conference on Digital Information and Communication Technology and its Applications (DICTAP)}, 2015, pp. 178-182.
\bibitem{cache big data}
    M. A. Kader et al., ``Leveraging Big Data Analytics for Cache-Enabled Wireless Networks," in \emph{2015 IEEE Globecom Workshops (GC Wkshps)}, San Diego, CA, 2015, pp. 1-6.
\bibitem{caching limits}
    M. A. Maddah-Ali and U. Niesen, ``Fundamental Limits of Caching," \emph{IEEE Transactions on Information Theory}, vol. 60, no. 5, pp. 2856-2867, May 2014.

\bibitem{context Schilit}
    B. Schilit, N. Adams and R. Want, ``Context-Aware Computing Applications," in \emph{Mobile Computing Systems and Applications, 1994. WMCSA 1994. First Workshop on}, Santa Cruz, California, USA, 1994, pp. 85-90.
\bibitem{Context Yang}
    J. Guo, C. Yao and C. Yang, ``Proactive resource allocation planning with three-levels of context information," in \emph{2016 IEEE/CIC International Conference on Communications in China (ICCC)}, Chengdu, 2016, pp. 1-6.
\bibitem{future direct}
    D. Liu, B. Chen, C. Yang and A. F. Molisch, ``Caching at the wireless edge: design aspects, challenges, and future directions," \emph{IEEE Communications Magazine}, vol. 54, no. 9, pp. 22-28, Sep. 2016.
\bibitem{cache user association}
    F. Pantisano, M. Bennis, W. Saad and M. Debbah, ``Cache-aware user association in backhaul-constrained small cell networks," in \emph{2014 12th International Symposium on Modeling and Optimization in Mobile, Ad Hoc, and Wireless Networks (WiOpt)}, Hammamet, 2014, pp. 37-42.
\bibitem{user access fog}
    S. Yan, M. Peng and W. Wang, "User access mode selection in fog computing based radio access networks," 2016 IEEE International Conference on Communications (ICC), Kuala Lumpur, 2016, pp. 1-6.







	









\end{thebibliography}
\end{document}